# IUPAC-Induced Computational Approaches for Identifying Boosters of Small Biomolecule Functionality: A Case Study of Human Tyrosyl-DNA Phosphodiesterase 1 (TDP1) Inhibitors

Mariya L. Ivanova[1,*, ID], Nicola Russo[1, ID], Gueorgui Mihaylov[2, ID] and Konstantin Nikolic[1, ID]

Author affiliations: [1]School of Computing and Engineering, University of West London, London, UK
[2]Haleon, London, UK
*Corresponding author mariya.ivanova@uwl.ac.uk

## Abstract

This paper introduces several proof-of-concept (PoC) computational methods intended to offer biochemical researchers straightforward, time- and cost-effective strategies to accelerate their work. While Machine Learning (ML) models were developed, the study's central purpose was to explore approaches for the identification of desirable functional groups/fragments in small biomolecules regarding a specific functionality, which, in this case, was human tyrosyl-DNA phosphodiesterase 1 (TDP1) inhibition. This was achieved primarily by tokenising IUPAC names to generate features. Additionally, the applicability of the CID_SID ML model for predicting TDP1 activity was developed and explored. Since these computational approaches were not experimentally validated due to a lack of appropriate laboratory facilities, they are presented as open proposals for further laboratory investigation.

## Introduction

This work aims to develop computational approaches capable of predicting potential drug side effects in the development pipeline and generating novel insights for drug discovery. A major challenge in drug development underscores the critical need for computational tools: a report [1] shows that nine out of every ten drug candidates fail to reach FDA submission after entering trials. This inefficiency is compounded by the fact that developing a new drug requires an average of 12 years and costs over a billion US dollars [2]. To develop the proposed supportive computational approaches, the strategy focused on leveraging the vast experimental data within the PubChem repository, alongside the use of relatively simple chemical nomenclature for naming compounds and substances. The information encoded in the names of compounds, which are generated according to the International Union of Pure and Applied Chemistry (IUPAC) nomenclature [3], was investigated to determine how it could be utilized. These unique IUPAC names encode the chemical composition and structure of compounds, thereby ensuring clear and unambiguous communication among chemists worldwide. The naming variations inherent in the IUPAC nomenclature, including the designation of a Preferred IUPAC Name (PIN) for regulatory precision, do not compromise the integrity of this study. While PINs resolve ambiguity by standardising the order of precedence or the choice of parent structure, they are fundamentally built upon the same chemical structure as other valid systematic names. Consequently, the set of functional groups or chemical fragments remains identical regardless of the name chosen. The methodology's core strength lies in explicitly tokenising the IUPAC names to extract the presence or absence of these discrete structural fragments (e.g., 'phenyl' or 'imidazo'), successfully bypassing the ordering variations that define a PIN. Therefore, the resulting binary feature matrix (data frame) is a robust, direct representation of the molecule's composition, untouched by naming conventions.

IUPAC names have previously been leveraged for computational drug discovery applications by the large language models (LLMs) iupacGPT [4] and BioT5+ [5]. However, the two LLMs possess significant differences. The iupacGPT [4] approach is highly focused, treating IUPAC names as a chemist's natural language to directly design new compounds and predict their functionalities. In contrast, BioT5+ [5] uses IUPAC names merely as one feature within a vast, integrated dataset. This broadened data includes literature from PubMed and bioRxiv, molecular data like SELFIES from PubChem, and protein sequences from UniRef50, allowing the model to learn simultaneously across 15 different tasks and 21 benchmark datasets. A limitation common to both LLMs is the risk of "hallucination" [6, 7], which, in chemistry, translates to generating non-existent or fabricated information. That includes reporting incorrect properties, non-existing interactions, or invalid sequences and nomenclature (SMILES, SELFIES, IUPAC names, or protein sequences) [8]. Prediction of IUPAC names based on the International Chemical Identifier (InChI) of the chemicals was performed by implementing character-by-character tokenisation of the names. This Machine Learning (ML) model achieved 91% accuracy for organic molecules (excluding macrocycles) [9], which suggests the potential for vice-versa implementation using IUPAC-encoded data. A separate insightful study demonstrated that deep learning (DL) can successfully generate IUPAC names directly from Atomic Force Microscopy (AFM) images [10]. Extensive review of the publicly available literature confirms that the approach presented in this paper is novel and has not been reported to date. The paper integrates two methodologies that process IUPAC-derived data to produce lists of functional groups. These lists are designed to facilitate drug development by serving as a resource for human-driven medicinal chemistry insights.

To demonstrate the methodologies, the PubChem AID 686978 bioassay was selected [11]. This assay is focused on human tyrosyl-DNA phosphodiesterase 1 (TDP1) and seeks to identify active inhibitors, thereby providing potential agents to modulate the TDP1-mediated repair pathway for cancer therapy. While not strictly an essential protein, TDP1 becomes critical for cell survival under treatment with the topoisomerase I poison camptothecin (CPT). To screen for inhibitors active in a cellular environment, a specialised assay was developed using chicken DT40 cells: a TDP1 knockout line (Tdp1 -/-) and a complemented line (Tdp1 -/-; hTDP1) stably expressing human TDP1. In the primary screen (PubChem AID 686978), the latter cells were exposed to small molecules from the MLSMR, both in the presence and absence of CPT, and their growth kinetics were evaluated by measuring ATP activity after 48 hours. A compound demonstrating a synergistic effect with CPT suggests inhibition of the CPT-induced repair pathway, potentially through TDP1. Compounds that showed synergy in Tdp1 -/-; hTDP1 cells but not in Tdp1 -/- cells were classified as hits potentially involved in TDP1-mediated repair inhibition and were then subjected to tertiary biochemical gel-based assays for specific TDP1 targeting assessment. The screening involved dispensing 400 DT40-hTDP1 cells per well into 1536-well plates, transferring 23 nL of compounds, incubating for 48 hr, and reading luminescence after adding Cell Titer Glo solution. Finally, compounds were ranked based on their titration curves, with active compounds assigned a PUBCHEM_ACTIVITY_SCORE between 40 and 100 (with Fit_LogAC50 used for relative scaling), inconclusive compounds scored 1 to 39, and inactive compounds scored 0 [11]. The enzyme TDP1 was selected as the demonstration object for this study because of its crucial function in the DNA repair pathway, specifically in resolving lesions caused by topoisomerase I cleavage complexes (TopIcc) [12].

The TDP1 enzyme's DNA repair capability holds promise for cancer treatment, as demonstrated by its clinical testing in a study involving 150 patients with non-small-cell lung cancer [13]. To elucidate the mechanism by which TDP1 repairs DNA-protein crosslinks (DPCs), which are DNA lesions leading to genomic instability and cell death, an investigation was performed that confirmed the endogenous role of TDP1 in DPC repair [14, 15].

In addition to its role in cancer, TDP1 has non-cancer related functions; specifically, its mutation is the known cause of spinocerebellar ataxia with axonal neuropathy type 1 (SCAN1) [16, 17]. This rare neurodegenerative disorder remains incurable to date, representing a critical gap in current medical research [18, 19].

An exploration of the available literature revealed a variety of AI approaches currently assisting in the field of drug discovery [20, 21]; feature engineering for ML models using atomic properties. [22]; the results of ten ML algorithms were compared to facilitate the prediction of the intervention age that would improve the efficacy of the treatment for spinocerebellar ataxia type 3 [23]; ML predictions based on 13C NMR spectroscopic data derived from Simplified Molecular-Input Line-Entry System (SMILES) [24]; an ML model developed to predict potential TDP1 inhibitors using SMILES notations [25] that were transformed into numerical data by the RDKit cheminformatics toolkit [26]. To date, the approaches based on IUPAC-parsed names presented in this article this article has not been reported in the literature.

While this work centres on approaches using parsed IUPAC names, the TDP1 inhibition data were also utilized to test the applicability of the CID_SID ML model methodology [27]. This secondary model, while also predicting TDP1 inhibition, has a distinct goal: it relies solely on the PubChem compound and substance identifiers (CID/SID) to computationally screen compounds originally designed for other purposes. While sample identifiers typically lack data suitable for ML training and testing, the PubChem compound and substance Identifiers (CID and SID) may be an exception. Since CID and SID are not arbitrary identifiers but sophisticated, structured keys to vast chemical and biological information, they can be used in ML. The CID acts as a non-redundant identifier for a single, canonical chemical structure; it is the result of a rigorous standardisation algorithm that resolves ambiguities like tautomers and salts, allowing it to serve as a reliable anchor for all molecular properties (e.g., fingerprints, SMILES strings) that are the true features in cheminformatics ML [28]. Conversely, the SID is crucial because it links the canonical structure (CID) back to the specific experimental context and data source (depositor) that provided the information, which is vital for modelling data variability, source bias, and linking to raw bioassay results. By using these IDs, the ML model implicitly taps into PubChem's extensive data hierarchy and curation. This allows the model to group similar compounds and contextualise activity data, a strategy that has proven successful in numerous predictive modelling studies. Capitalising on this, a study developed CID_SID ML models, predicting D3 dopamine receptor antagonists, Rab9 promoter activators, DNA damage-inducible transcript 3 inhibitors and M1 muscarinic receptor antagonists [27]. The general applicability of the CID_SID ML model was later confirmed by a separate study that successfully predicted dopamine D1 receptor antagonists [24]. Leveraging this proven methodology, a CID_SID ML model was subsequently developed in the current study to specifically predict TDP1 inhibitors.

## Methodology

The first methodology involves an ML model that uses IUPAC-tokenised data to predict a small biomolecule's functionality [29]. Subsequently, the feature importance from this model is extracted and processed to derive insights for biochemical and medical research. The second methodology also utilizes tokenized IUPAC names but focuses on generating ranked lists of functional groups. This ranking identifies the most and least desirable functional groups based on their correspondence with relevant labels across the entire High-Throughput Screening (HTS) bioassay dataset. The third methodology is a complementary ML model that replicates the previously established CID_SID ML model. This model does not rely on IUPAC names but serves as a resource for researchers interested in screening compounds designed for other purposes to determine if they also possess TDP1 inhibitory capabilities.

**The first methodology** is illustrated in Figure1. As was noted above, for the purpose of demonstrating the proposed methodology, the case study was built upon the publicly available PubChem AID 686978 bioassay dataset focused on TDP1 inhibition [11]. The critical importance of data quality and the valid point regarding the need to comment on curation, even for well-known datasets, are acknowledged, as model validity can still be compromised by errors and mislabeled entries. Therefore, to ensure the reliability of the results, a rigorous data curation process was implemented prior to model training. This included the removal of duplicate entries, the conducting of consistency checks on all chemical structures and their corresponding activity labels, and the performing of literature cross-validation (CV) to resolve any ambiguous assignments. Through this thorough validation step, it is ensured that the final performance of the ML models, including the RFC, is based on a clean, statistically reliable foundation, thus addressing the potential for data-related inconsistencies.

The PubChem AID 686978 bioassay dataset consists of 424,883 samples (rows) defined by 48 features (columns). The samples were divided into three distinct activity groups: 64,192 active, 116,652 inconclusive, and 243,131 inactive. For the purpose of the first and second methodology, the active and inactive compounds were gathered into a single dataset. To mitigate the severe imbalance between the active and inactive small biomolecules, this resulting dataset was merged with the PubChem AID 1996 bioassay dataset [30], and only the common samples for both bioassays were retained.

While the filtering strategy using PubChem AID 1996 bioassay[30], focused on the aqueous solubility of small biomolecules, introduces a selection bias, this bias was considered intentional and necessary to focus the model on the chemical space most relevant to successful drug discovery; the bias was directed toward compounds with reliable physicochemical properties. Although the PubChem AID 1996 bioassay dataset contained 57,859 rows of samples, 40,860 of which were labelled as soluble and 17,573 as insoluble small biomolecules, solubility was not taken into consideration for the purpose of the current study. Only the column with CIDs was used to reduce the inactive compounds in the PubChem AID 686978 bioassay, keeping the common for both bioassay samples. Furthermore, the final dataset used for downloading of IUPAC names from PubChem was obtained by the addition of all active compounds to the reduced inactive compounds. In this way, simultaneous mitigation of both data imbalance and selection bias was achieved.

PubChem's requirement for retrieving a bulk query of IUPAC names is to specify a list containing only CIDs. After such a list was created from the resulting dataset, the file with IUPACs was downloaded via the PubChem home page,  Upload ID List, which is a data retrieval option for an easy-to-use way to perform bulk queries on the database without the need for more complex programming tools. For the purpose of the study, the comprehensive downloaded file from PubChem was filtered to retain only the necessary columns: CID, the SMILES string, and the IUPAC name. This focused dataset was then merged with the primary target-containing dataset. The merge operation used a compound key consisting of both the CID and the SMILES string to ensure accurate and non-redundant matching of each chemical entity with its corresponding experimental activity data. Following this procedure, to identify key molecular features, the IUPAC names in the dataset underwent a parsing step. Only strings of four or more letters were retained, as these likely represented significant functional groups or molecular fragments. The approach was intentionally designed and executed to ensure that the extracted strings represented existing constituents of the IUPAC names, meaning the process was constrained to avoid generating novel strings by cutting across IUPAC-defined groups or molecular fragments. The resulting strings then formed the column headers of a new data frame (Table 1). For each compound, a binary indicator (1 for present, 0 for absent) was assigned to mark the presence or absence of a given functional group, Table1. The feature assignment is based strictly on an absolute, exact string match between a functional group or fragment and the corresponding column (feature) name. A partial or subordinate match is not sufficient for assignment; for instance, the mere presence of the string "amino" within a molecule will not assign a label of "1" to every column name that contains "amino"; the functional group's string must perfectly equal the feature column's name. The new data frame, containing the functional group information, was subsequently merged with the labelled data frame by matching their CIDs and SMILES.

ML conducted in this study centred on the Random Forest Classifier (RFC), whose interpretation was facilitated by the scikit-learn library. The entire ML process, encompassing data preprocessing, model training, prediction, and result evaluation using relevant metrics, was compiled and executed in accordance with the best relevant practices recommended in the literature, as detailed in references [31], [32], and [33].

After the tokenisation of IUPAC names and integration with their relevant labels, the dataset was split into data points (X) and targets (y). These were subsequently divided into training and test sets (X_train, X_test, y_train and y_test). The test set was manually created by randomly extracting an equal number of cases to ensure a balanced evaluation. Using an equal number of classes in the test set (a balanced test set of randomly selected samples) basically provided a fair and reliable evaluation of the ML model's ability to generalise across all outcomes. This approach directly tackles the accuracy paradox, where an imbalanced test set could yield a deceptively high accuracy score by a model that simply predicts the majority class, masking poor performance on the minority class. By employing equal class proportions, the model is forced to correctly identify instances from every class, ensuring the overall accuracy is a meaningful metric that reflects equal emphasis and penalty for misclassification across all outcomes. Furthermore, a balanced test set allows the straightforward and reliable interpretation of standard metrics such as the Precision, the Recall and the F1-Score, validating the model's robustness and confirming that it has learned the distinct patterns for each classification label rather than merely exploiting a data distribution bias. The remaining samples formed the training set. It was then balanced using Random Over Sampler (ROS), which randomly replicated minority class samples until a balanced training dataset was achieved. Finally, these prepared sets were used for training, predicting, and evaluating the ML model based on the Random Forest Classifier (RFC) strategy. To ensure the ML model's reported performance (including Accuracy, Precision, Recall, F1-score, and ROC) is robust and generalised, rather than relying on a single data split, five-fold CV was conducted. The procedure began by partitioning the shuffled dataset into five equal, stratified folds to preserve the class proportion. In each iteration, one fold was designated as the test set, while the remaining four were used for training. This training set was further processed using the balancing technique appropriate for the model. The model was trained iteratively on the four balanced folds and validated on the single, untouched test fold, with all metrics recorded. The model's final, generalised performance is reported as the mean and standard deviation of the five recorded scores, confirming consistency.

Table 1. Methodology. Parsing/breaking down the IUPAC names into tokens/strings equal or longer than four letters and using these tokens/strings to create the data frame features. Counting the presence of the functional groups in the compound`s content with 1 for presence and 0 for absence of the relevant feature group/fraction in the content of the small biomolecule.

| IUPAC | amino | oxopropyl | butan | methylpropyl | methylundecyl | octaoxo | heptazacyclopentaco | acetic | acid | pent | enedioic | cyclopropylsulfonylph | oxan | yloxy | acetyl | methoxybenzoic | thiazol | trimethyl | methylpiperidin | … | oxyanilino | oxohexan | dihydro | indole | carboxamide |
|---|---|---|---|---|---|---|---|---|---|---|---|---|---|---|---|---|---|---|---|---|---|---|---|---|---|
| 2-[21-(3-amino-3-oxopropyl)-3,12,18-tri(butan-2-yl)-6,15-bis(2-methylpropyl)-25-(9-methylundecyl)-2,5,8,11,14,17,20,23-octaoxo-1-oxa-4,7,10,13,16,19,22-heptazacyclopentacos-9-yl] acetic acid | 1 | 1 | 1 | 1 | 1 | 1 | 1 | 1 | 1 | 0 | 0 | 0 | 0 | 0 | 0 | 0 | 0 | 0 | 0 | … | 0 | 0 | 0 | 0 | 0 |
| (*E*)-pent-2-enedioic acid | 0 | 0 | 0 | 0 | 0 | 0 | 0 | 0 | 1 | 1 | 1 | 0 | 0 | 0 | 0 | 0 | 0 | 0 | 0 | … | 0 | 0 | 0 | 0 | 0 |
| 4-[[2-[[2-(4-cyclopropylsulfonylphenyl)-2-(oxan-4-yloxy)acetyl]amino]-1,3-thiazol-5-yl]oxy]-2-methoxybenzoic acid | 1 | 0 | 0 | 0 | 0 | 0 | 0 | 0 | 1 | 0 | 0 | 1 | 1 | 1 | 1 | 1 | 1 | 0 | 0 | … | 0 | 0 | 0 | 0 | 0 |
| ⋮ | ⋮ | ⋮ | ⋮ | ⋮ | ⋮ | ⋮ | ⋮ | ⋮ | ⋮ | ⋮ | ⋮ | ⋮ | ⋮ | ⋮ | ⋮ | ⋮ | ⋮ | ⋮ | ⋮ | ⋮ | ⋮ | ⋮ | ⋮ | ⋮ | ⋮ |
| 3,6,6- -*N*-[(2*S*)-1-[4-(1-methylpiperidin-4-yl)oxyanilino]-1-oxohexan-2-yl]-4-oxo-5,7-dihydro-1*H*-indole-2-carboxamide | 0 | 0 | 0 | 0 | 0 | 0 | 0 | 0 | 0 | 0 | 0 | 0 | 0 | 0 | 0 | 0 | 0 | 1 | 1 | … | 1 | 1 | 1 | 1 | 1 |

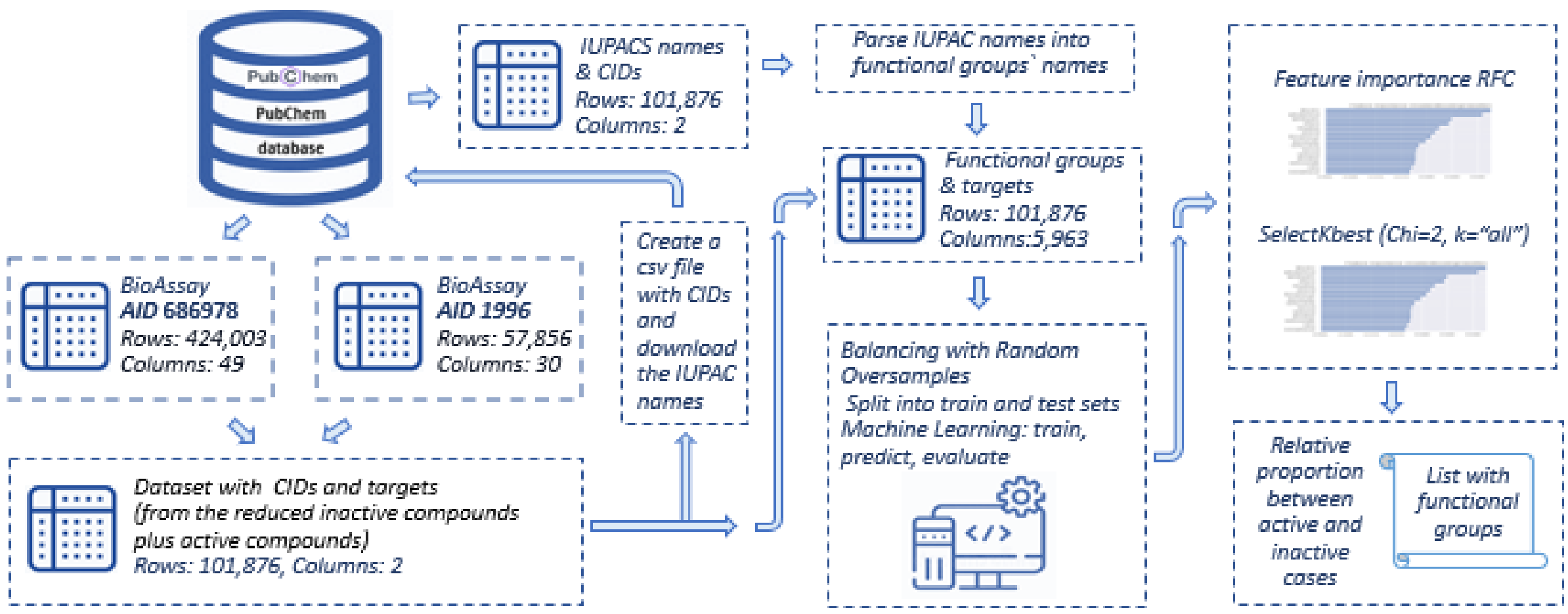

Figure 1. Methodology of the ML model utilizing
IUPAC names, PubChem bioassay data and the scikit learn ML library

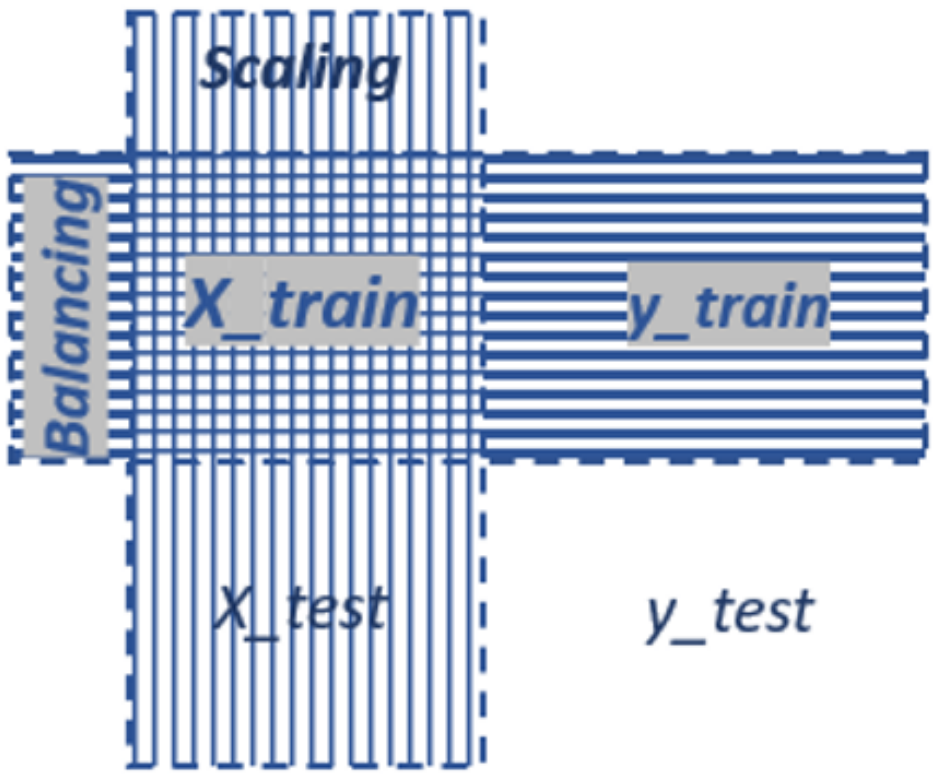

Figure 2.
Illustration of Data Preprocessing Steps
(Balancing and Scaling) Applied to the Training and Test Datasets.
(Ivanova et.al, 2025)

Scientific context and proof of performance were established by comparing the model against two industry-standard baselines: MORGAN2 fingerprints [34] and features computed by RDKit from the SMILES notation [25]. These RDKit representations were essential as they provide a proven, accurate, and highly efficient means of translating complex molecular information into the fixed numerical format that ML and QSAR algorithms can process.

Principal Component Analysis (PCA) was implemented as an unsupervised dimensionality reduction technique [35]. Its core function was to simplify the complex, high-dimensional dataset by transforming the original, correlated variables into a smaller, more manageable set of uncorrelated variables known as Principal Components (PCs), thereby retaining the maximum possible data variability and essential information

Two complementary feature inspection techniques, both implemented via Scikit-learn, were used to rank the functional groups that compose the small molecule TDP1 inhibitors. The primary ranking was established using the RFC feature importance, which sorted the functional groups in descending order based on their ability to reduce Gini impurity, quantifying the model's reliance on each group for accurate prediction. A second, independent ranking was created using the Chi-squared statistical test within the SelectKBest tool. This test assesses the statistical independence of each functional group from the target variable; since highly independent variables have low predictive value, this process effectively quantifies the potential predictive power of each group by measuring the strength of its direct relationship with the inhibition data. Collectively, these methods provided a clear quantification of feature influence by systematically assessing and disrupting the feature-target relationship.

After generating two distinct feature importance lists from the RFC model, each likely derived from a different set of molecular representations, a single, consolidated list of unique functional groups was created. This final, unified list was then re-ranked using two independent chemical relevance metrics. The first re-ranking was based on the relative proportion of active versus inactive small biomolecules that possessed the given functional group, directly correlating group presence with target activity (as detailed in Table 2). The second, more rigorous re-ranking utilized Fisher's exact test [36] to statistically assess the non-random association (dependency) between the categorical variable (the presence of a functional group) and the binary target outcome (active or inactive), providing a precise measure of the group's statistical significance as a predictor (Table 2). The Boruta feature selection algorithm [37] was additionally employed to identify the most relevant chemical features, providing a statistically robust foundation for subsequent process validation. By ensuring the ML models (such as the RFC) were trained only on the most significant variables, Boruta allowed for a rigorous and fair comparison of their performance, thereby confirming that the final results were based on truly meaningful inputs.

To optimise the model's performance, hyperparameter tuning was conducted on the RFC using Bayesian optimisation. This entire process was automated and executed efficiently via the open-source framework, Optuna [38]. During the hyperparameter optimisation using Optuna, the process was guided by k-fold Stratified CV (StratifiedKFold) on the training data (X_train and y_train), where the objective function iteratively tests hyperparameter combinations and returns the mean cros_val_score to maximise performance while preventing overfitting to any single data subset. The X_train set was fully utilized for both training and validation via the 5-fold CV to provide a stable, low-variance estimate of the optimal parameters. Finally, once the best parameters were found, the model was refit on the entire training set (X_train, y_train), and it was true, unbiased generalisation performance assessed only once by calculating the scikit-learn method .score() on the completely held-out test set (X_test and y_test), ensuring the reported result was a realistic measure of performance on unseen data. X_test and y_test were not used in any way to influence the model's structure, weights, or hyperparameters during the crucial learning and tuning phases.

Tracing the deviation between the training accuracy (the model's performance on seen classification examples) and the test accuracy (its ability to correctly classify unseen examples) is the most direct way to detect and quantify overfitting. Ideally, both curves increase together, confirming the model is learning generalisable classification boundaries. However, in an overfit scenario, the curves diverge: the training accuracy continues its ascent as the model perfectly memorises the training data's noise and specific class assignments, while the test accuracy plateaus and then drops because the over-specialized model fails to generalise to new data. The resulting size of the gap between these two metrics directly indicates the level of overfitting, making this trace critical for implementing early stopping, a necessary regularisation technique that halts training when test accuracy starts its decline, thereby selecting the classification model with optimal real-world generalisation capability. Although setting a fixed 5% deviation threshold for classifying 'overfitting' is arbitrary, in this specific context, the 5% margin was used only as an initial threshold to monitor model stability during preliminary runs. The primary and standard criterion for mitigating and defining overfitting was the use of early stopping, where training was halted precisely when the test (validation) loss failed to decrease for a set number of epochs. Therefore, the 5% threshold served merely as a secondary, conservative monitoring tool, and the true measure of acceptable generalisation was based on the F1-score and ROC metrics.

The Matthews Correlation Coefficient (MCC) was used as a single, robust metric to assess the quality of a binary classification model, particularly because it provides a reliable score even when dealing with imbalanced datasets [39]. The MCC is a correlation coefficient between the true and predicted classifications, symmetrically incorporating all four confusion matrix components: True Positives (TP), True Negatives (TN), False Positives (FP) and False Negatives (FN), into its calculation, ensuring that a high score reflects strong performance across all classes, not just the majority one. A perfect model scores +1, a random guess scores 0, and a perfect inverse prediction scores -1. However, MCC's limitations include its complex formula, which makes the magnitude less intuitive than Accuracy, its sensitivity to the classification threshold, and the fact that its formula can be undefined in rare, degenerate cases where a row or column in the confusion matrix is empty.

Confidence Intervals (CIs) were calculated for the ML parameters to quantify the uncertainty and precision associated with the sample-based estimates, thereby establishing a plausible range of values for the true population parameters. This process fundamentally acknowledges that any estimate derived from a limited sample is inherently subject to random error, with the resulting CI's width serving as a direct indicator of the estimate's precision. Furthermore, CIs are crucial for statistical significance testing, as they enable researchers to quickly ascertain if an effect is significant (e.g., by observing if the interval for a coefficient excludes zero) or if the performance of different models is meaningfully distinct (by assessing the overlap of their respective CIs). Ultimately, the inclusion of CIs enhances scientific reporting by lending transparency, context, and reliability to the reported results.

Local Interpretable Model-agnostic Explanations (LIME) was used to provide essential transparency for any "black-box" ML model, which is critical for trust, debugging, and compliance. By creating a simple, local explanation for individual predictions, LIME allowed users to verify that the model was making decisions based on sound, expected features, rather than spurious correlations or data leakage. This capability is paramount in high-stakes fields to diagnose flaws before deployment and is increasingly necessary to meet regulatory and ethical requirements for model accountability and fairness.

**The second methodology** aimed to identify chemical groups exclusively associated with a single activity class by focusing on the chemical composition of small biomolecules exhibiting a purely active or inactive nature regarding TDP1 inhibition (as summarised in Figure 3). To achieve this, all molecules from AID 686978 were first tokenised and organised into a data frame containing their IUPAC names and corresponding target labels (active/inactive), following the same pre-processing steps as the first methodology. The core step then involved selectively extracting functional groups that appeared only in the composition of active biomolecules and were entirely absent from inactive ones. This process was then repeated to isolate functional groups found only in inactive biomolecules, thereby generating two distinct lists of activity-exclusive chemical features.

For both methodologies (First and Second), the compounds containing the suggested functional groups were screened for Pan-Assay Interference Compounds (PAINs) [40]. Despite their known propensity for generating false positives and assay artefacts they were not removed in advance. Fundamentally, PAINS represent statistical alerts rather than absolute exclusion rules, as evidenced by the small percentage of FDA-approved drugs that successfully contain these motifs [41]. Strictly filtering all PAINS risks incurring detrimental false negatives, leading to the premature dismissal of genuinely active or unique scaffold hits simply because they share a substructure common to interferers. Furthermore, retaining these compounds preserves invaluable historical data from legacy HTS campaigns, which is essential for a comprehensive chemical context. Finally, from an ML perspective, including flagged PAINS is necessary to train robust models that can effectively learn and predict promiscuity and chemical reactivity, ultimately enhancing the model's ability to triage novel compounds based on a complete spectrum of chemical behaviour. Therefore, their presence in large datasets is not accidental but serves to preserve historical integrity, prevent the loss of potential leads, and improve predictive modelling power.

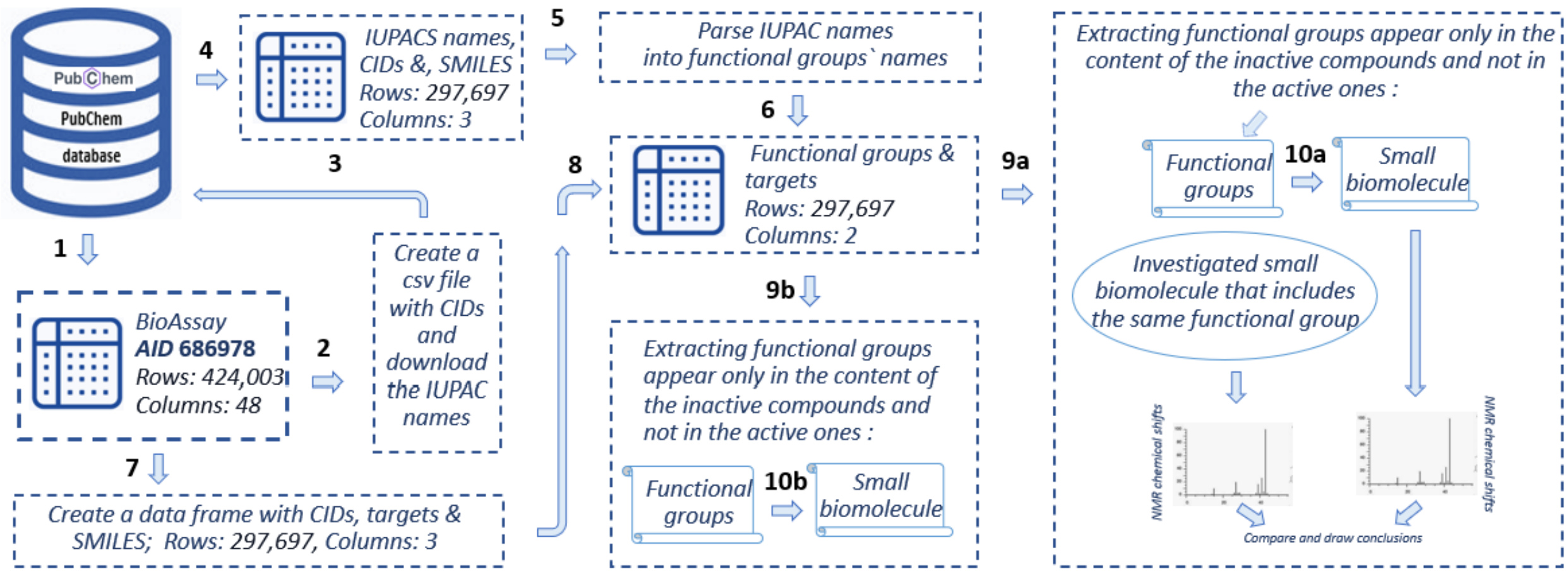

Figure 3. Methodology generating ranking of the functional groups according to their presence in the biomolecule content

**The third methodology**, complementary to the main study, adapted the framework of established CID-SID ML models to develop a dedicated TDP1 CID-SID model [27]. The data was sourced from PubChem AID 686978 [11], retaining only the CID, SID, and the column related to TDP1 activity. The imbalance of the resulting dataset was addressed by initial filtering with the PubChem AID 1996 bioassay [30], followed by the described oversampling technique. The model leveraged five diverse classifiers: Decision Tree Classifier (DTC), RFC, Gradient Boosting Classifier (GBC), Extreme Boosting Classifier (XGBC), and Support Vector Classifier (SVC). To maximise predictive capability, the models underwent rigorous evaluation, including five-fold CV, hyperparameter optimisation with Optuna [37] and overfitting analysis as explained above. The performance of the resulting optimal model was then benchmarked against the MORGAN2 and SMILES RDKit-based ML models established in the third methodologies. PCA and feature importance were not processed because of the volume and nature of the datapoints columns (i.e. two with identifiers). Finally, MCC and confidence intervals for key ML metrics (Accuracy, Precision and Recall) were calculated across all models to quantify the uncertainty of the performance estimates.

## Results and Discussion

### Results regarding the first methodology

After removing the isomers without retaining any samples, the dataset was reduced to 61,471 active, 112,867 inconclusive, and 236,226 inactive compounds. This reduced set was then filtered using the PubChem AID 1996 bioassay [30], which left 40,404 inactive compounds. Concatenating these remaining inactive compounds with the active compounds (minus the isomers) yielded the final dataset of 101,860 samples [42]. The IUPAC names for these samples were downloaded from PubChem and parsed into strings of four or more letters, a process that generated 5,963 features (columns) for the model. The training of the RFC on 64,625 samples and testing on 28,200 samples resulted in a strong initial performance. The model demonstrated a good balance of overall correctness, reflected by an accuracy of 78.4%, and a high confidence in its positive predictions, indicated by a precision of 83.4%. While the recall was lower at 70.8%, suggesting a notable portion of actual positive cases were missed, the combined metric of the F1-score (76.6%) confirms a solid harmonic mean between precision and recall. Finally, the ROC of 78.4% indicates the model has a promising ability to distinguish between the classes. This performance establishes a robust and encouraging baseline for the classification task. This achievement was further substantiated by five-fold CV(Table 2), which yielded a validated mean accuracy of 75.45% ±0.49 and a significantly high mean precision of 89.71% ±0.27, though a lower mean recall of 69.82±0.74 was also observed. Crucially, a robust and stable performance was confirmed by a strong mean F1-score of 78.52±0.51 and a high mean ROC of 85.80±0.41, with the tight standard deviations indicating that model consistency was successfully maintained across the data partitions.

Table 2. Five-fold cross-validation with StratifiedKFold of RFC based on IUPAC tokenised data

| Metrics | | Fold | | | | | Average across the folds |
|---|---|---|---|---|---|---|---|
| | | 1 | 2 | 3 | 4 | 5 | |
| Accuracy [%] | | 75.28 | 74.88 | 76.24 | 75.08 | 75.75 | 75.45±0.49 |
| Precision [%] | | 89.63 | 89.23 | 89.79 | 89.85 | 90.04 | 89.71±0.27 |
| Recall [%] | | 69.60 | 69.29 | 71.13 | 69.04 | 70.04 | 69.82±0.74 |
| F1-Score [%] | | 78.35 | 78.00 | 79.38 | 78.08 | 78.79 | 78.52±0.51 |
| ROC AUC [%] | | 85.64 | 85.24 | 86.37 | 85.58 | 86.19 | 85.80±0.41 |
| Confusion Matrix | TN | 4498 | 4469 | 4495 | 4522 | 4526 | 4502.0±23.08 |
| | FP | 763 | 792 | 766 | 739 | 734 | 758.8±23.34 |
| | FN | 2879 | 2909 | 2737 | 2932 | 2838 | 2859.0±76.74 |
| | TP | 6592 | 6562 | 6737 | 6539 | 6634 | 6612.8±78.00 |

The model's optimal performance, based purely on test accuracy before significant deviation, was observed at max_depth=17, where the test Accuracy was 69.8% and the deviation between test and train accuracy was still below 5%. This setting is likely the most generalisable because when max_depth was increased to 19, a clear sign of overfitting was encountered: the train Accuracy jumped to 75.5% while the test Accuracy dropped slightly to 70.3%. Crucially, the deviation between the two exceeded 5% for the first time, indicating that the model began to fit the training noise rather than the underlying pattern, losing its ability to generalise (Figure 4).

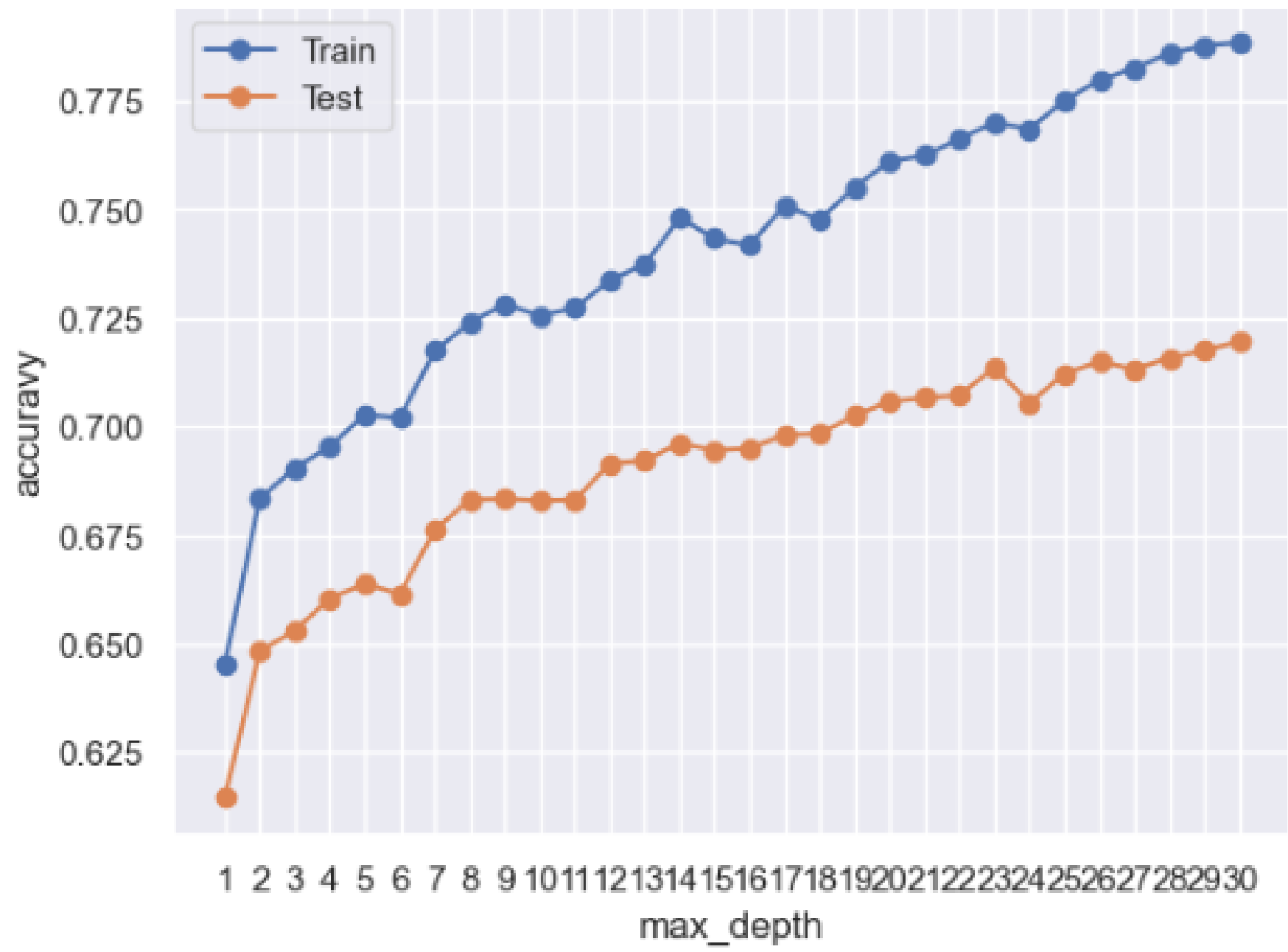

Figure 4. Overfitting analysis of the IUPAC RFC ML model: prediction accuracy vs maximum depth of the decision tree. The blue line is the training accuracy. The orange line is the test accuracy. The deviation between the testing and training accuracy higher than 5% was considered as an indication of overfitting.

The implementation of PCA reduced the number of features from 5, 963 to 44. However, the ML model performed with the PCA reduction of the features obtained accuracy 69.7%, precision 65.8%, recall 82.1, F1 73%, ROC 69.7% which metrics values were a bit lower compared to these obtained by the ML model without PCA reduction of the features.

The Optuna hyperparameter search was highly effective, yielding an optimal Random Forest configuration defined by: 'n_estimators': 404, 'max_depth': 8, 'min_samples_split': 10, 'min_samples_leaf': 6, 'max_features': 'log2', 'criterion': 'gini'. This finely-tuned model achieved a test Accuracy of 0.7068, successfully outperforming the 69.4% accuracy baseline obtained with default features and max_depth of 17. Crucially, the substantial reduction in max_depth from 17 to 8 and the tuning of the splitting criteria parameters demonstrate a clear strategy to combat overfitting, which was confirmed by the model adhering to the strict 5% deviation threshold between train and test sets, validating the Optuna result as the optimal balance of performance and generalisation. All key metrics cluster around 70.7%, indicating a good overall balance in the model's predictive capability. Specifically, the Precision is slightly higher at 0.7141 than the Recall at 0.691, suggesting the model is marginally better at making correct positive predictions than at identifying all actual positive cases. However, tracing the deviation between the train and test accuracy of the RFC based on Optuna hyperparameters revealed that even at max_depth=30, the difference remained under 5%. Consequently, the max_depth was set to its default value, None. This allowed the trees to continue growing until every leaf was pure (contained only samples of one class) or until a leaf contained fewer than min\samples\split samples.

The Optuna hyperparameter-tuned RFC with max_depth=None demonstrated moderate and highly stable performance on the testing data, achieving an Accuracy of 0.7233. All major metrics, Accuracy, Precision (0.7203), and Recall (0.7299), are tightly clustered just above the 70% benchmark, indicating the model was reliably better than random chance and maintains a good balance between FP and FN. Crucially, the narrow 95% CI, such as 0.7180 to 0.7283 for Accuracy, are the most significant finding. These narrow ranges, derived from 1000 bootstraps, confirm the model's exceptional robustness and consistency, suggesting that while the performance level is only moderate, it will not significantly degrade when applied to new, similar data.

RFC exhibits moderate predictive strength with an MCC of 0.4466, placing its overall performance significantly above random chance but well below perfect classification. The model shows a balanced distribution of errors, as indicated by the high and nearly equal counts of FP (3,996) and FN (3,808). While the model correctly classified a large number of instances (TP: 10,292 and TN: 10,104), this large volume of both types of misclassification suggests the model struggles to generalise effectively to the underlying decision boundary (Figure 5). Consequently, any future improvement efforts must focus on strategies to simultaneously suppress both the FP and FN rates to achieve a more robust MCC score closer to 1.

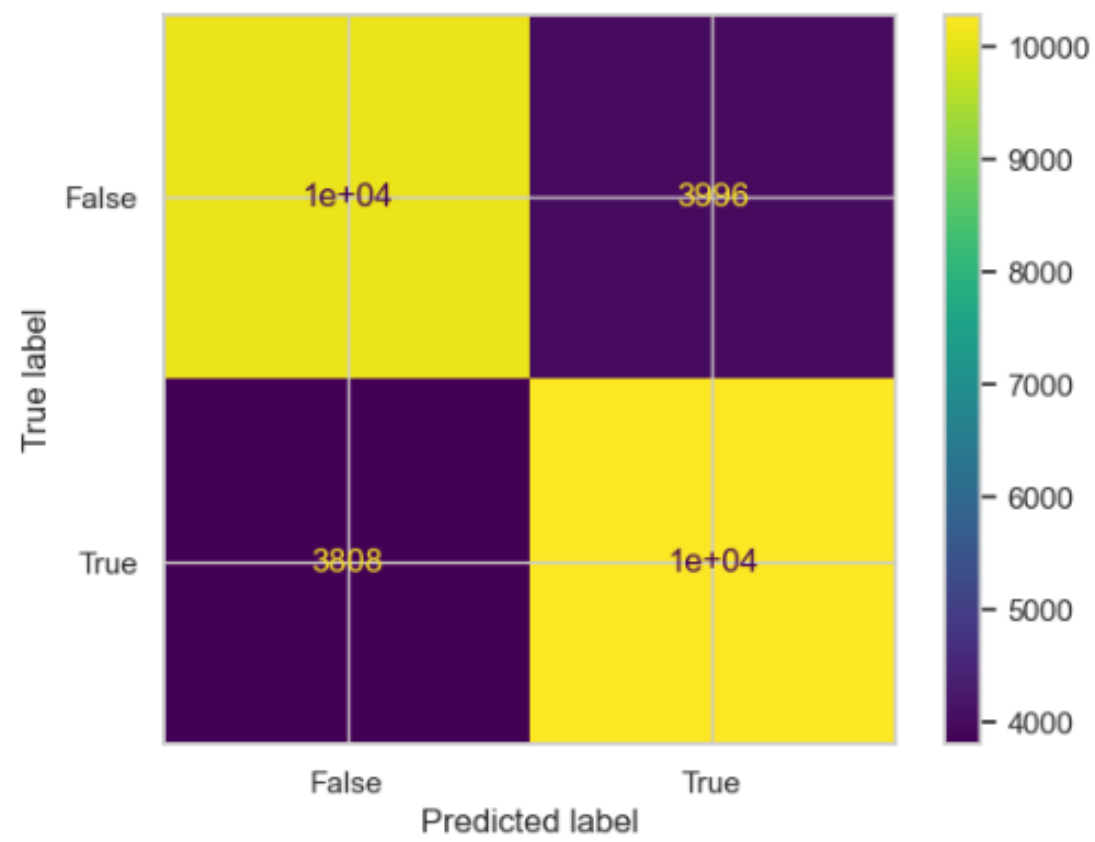

Figure 5.
The Confusion matrix of RFC based on IUPAC tokenised data

The classification report for the RFC (Table 3) reveals moderate and exceptionally balanced performance across both classes, achieving an overall Accuracy of 0.72 on a perfectly balanced test set of 28,200 samples. The most striking observation is the remarkable consistency: all key metrics, Precision, Recall, and F1-score, for both the Active (Target 1) and Inactive (Target 0) classes are tightly grouped between 0.72 and 0.73. This uniformity, confirmed by macro and weighted averages of 0.72, confirms the model is unbiased and generalises consistently across both outcomes. While its stability is a strength, the moderate 72% performance ceiling suggests the model may have reached its limit given the current features, and further accuracy improvements will likely require feature engineering or exploring more complex algorithms.

Table 3 Classification report of the IUPAC RFC ML model

| | precision | recall | F1-score | support |
|---|---|---|---|---|
| Active (target 1) | 0.73 | 0.73 | 0.72 | 14100 |
| Inactive (target 0) | 0.72 | 0.73 | 0.73 | 14100 |
| | | | | |
| accuracy | | | 0.72 | 28200 |
| macro avg | 0.72 | 0.72 | 0.72 | 28200 |
| Weighted avg | 0.72 | 0.72 | 0.72 | 28200 |

The provided LIME result for RFC based on IUPAC tokenised data pertains to a correctly classified instance, where both the True Label (1) and the Model Prediction (1) align, within a high-dimensional feature space (94,712 samples and 5,961 features) and a moderate overall Model Accuracy of 0.7235. The explanation identifies the two most influential features driving this specific prediction. Crucially, the absence or negligible presence (indicated by ≤0.00) of the chemical fragments 'phenoxythieno' and 'methoxyxanthen' strongly contributed to the model's decision. The negative LIME weights (-0.0204 and -0.0087) associated with the ≤0.00 condition mean that the presence of these features would have pushed the prediction away from class 1. Therefore, their absence provided the necessary local support, acting as a positive indicator for the correct classification of this instance as Label 1.

The RFC applied to the transformed MORGAN2 dataset obtained an Accuracy of 86.1%, a Precision of 89.1%, a Recall of 87.6%, an F1-score of 88.4% and an ROC of 85.7%. Tracking for the point where the deviation between train and test accuracy was higher than 5% point max_depth=10, where the train Accuracy was 79.7% and the test Accuracy was 74.6%. So, the choice was max_depth=9 with train Accuracy 78.7% and test accuracy 74.0%. This process, which guided the model's optimisation, is visually shown in Figure 6.

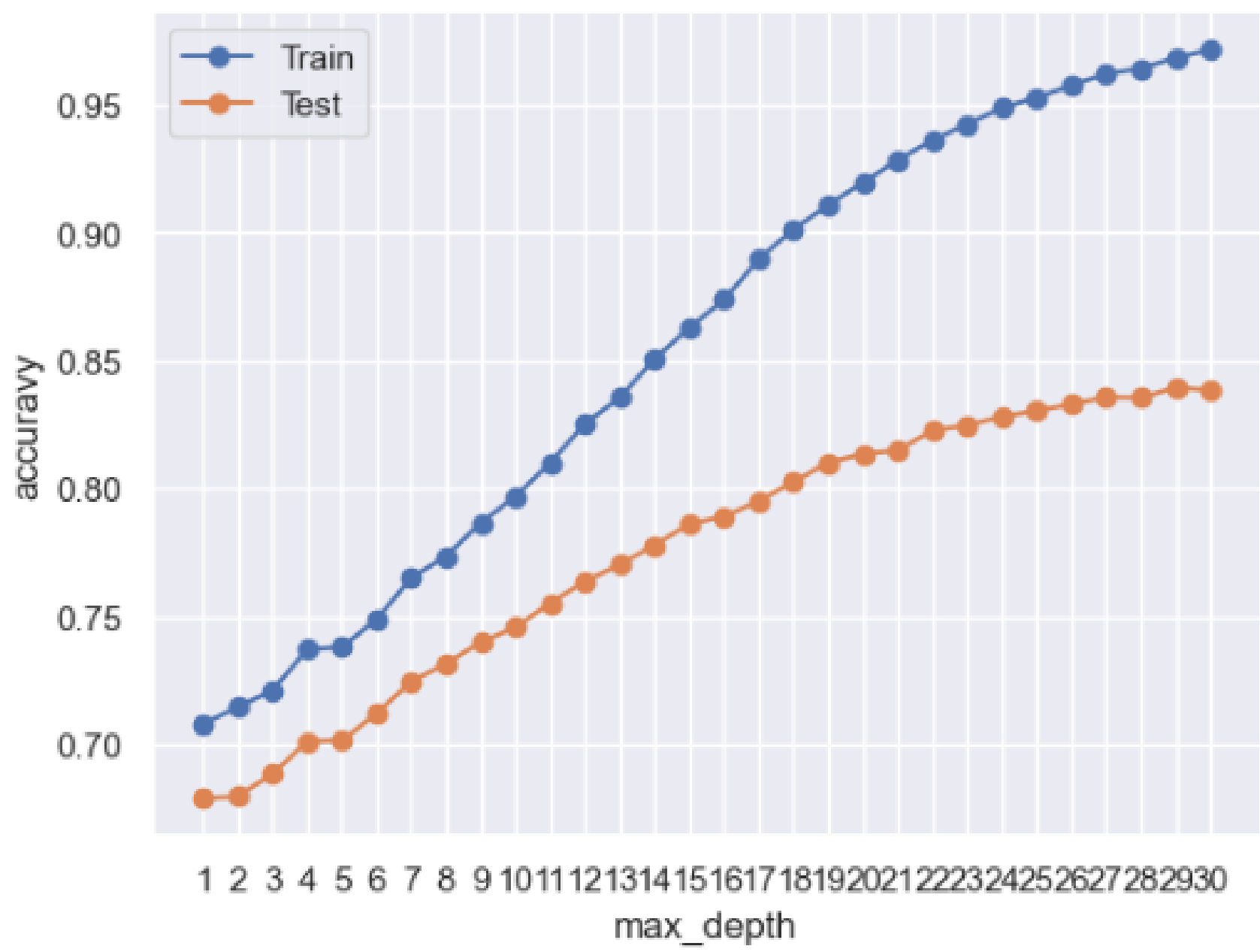

Figure 6.
Overfitting analysis of the RFC based on transformed MORGAN2 dataset: prediction accuracy vs maximum depth of the decision tree. The blue line is the training accuracy. The orange line is the test accuracy. The deviation between the testing and training accuracy higher than 5% was considered as an indication for early stopping.

Strong and stable performance was exhibited by the RFC based on the transformed MORGAN2 dataset when it was run with a maximum depth of 9. The model's single-run effectiveness was demonstrated by an Accuracy of 74.0%, a high Precision of 81.7%, a Recall of 73.3%, an F1-score of 77.3%, and an initial ROC of 74.2%. Crucially, the model's reliability and generalizability were confirmed through CV, where metrics were maintained with very low variance: the Accuracy was found to be 73.96% ±0.39, the Precision 81.44% ±0.33, the Recall 73.62%±0.43, the F1-score 77.33% ±0.36, and the ROC AUC was observed to be 81.58% ±0.37, which suggests the model possesses excellent discriminative ability that is highly consistent across different data subsets. The results are shown in detail in Table 4.

Table 4. Five-fold cross-validation of RFC based on MORGAN2 transformed features with StratifiedKFold

| **Metrics** | | **Fold** | | | | | **Average across the folds** |
|---|---|---|---|---|---|---|---|
| | | **1** | **2** | **3** | **4** | **5** | |
| **Accuracy [%]** | | 74.32 | 74.27 | 74.24 | 73.52 | 73.43 | 73.96±0.39 |
| **Precision [%]** | | 81.67 | 81.74 | 81.68 | 80.93 | 81.16 | 81.44±0.33 |
| **Recall [%]** | | 74.07 | 73.85 | 73.86 | 73.42 | 72.88 | 73.62±0.43 |
| **F1-Score [%]** | | 77.68 | 77.60 | 77.58 | 76.99 | 76.80 | 77.33±0.36 |
| **ROC AUC [%]** | | 81.95 | 81.94 | 81.55 | 81.52 | 80.95 | 81.58±0.37 |
| **Confusion Matrix** | **TN** | 4830 | 4843 | 4836 | 4763 | 4801 | 4814.6±29.49 |
| | **FP** | 1635 | 1622 | 1629 | 1701 | 1663 | 1650.0±29.05 |
| | **FN** | 2550 | 2571 | 2570 | 2614 | 2667 | 2594.4±41.88 |
| | **TP** | 7283 | 7262 | 7263 | 7219 | 7166 | 7238.6.2±41.88 |

The performance of the RFC on the transformed MORGAN2 dataset was quantified using the provided classification metrics, the details of which are visually represented in Figure 7. A total of 10,142 instances were accurately predicted as positive (TP), while 6,475 instances were correctly classified as negative (TN). Classification errors were noted, with 1,606 negative instances being incorrectly labelled as positive (FP) and 2,149 positive instances being missed and incorrectly labelled as negative (FN). These outcomes collectively resulted in an MCC of 0.6202, indicating a moderately strong and balanced measure of predictive quality across all four categories of the confusion matrix.

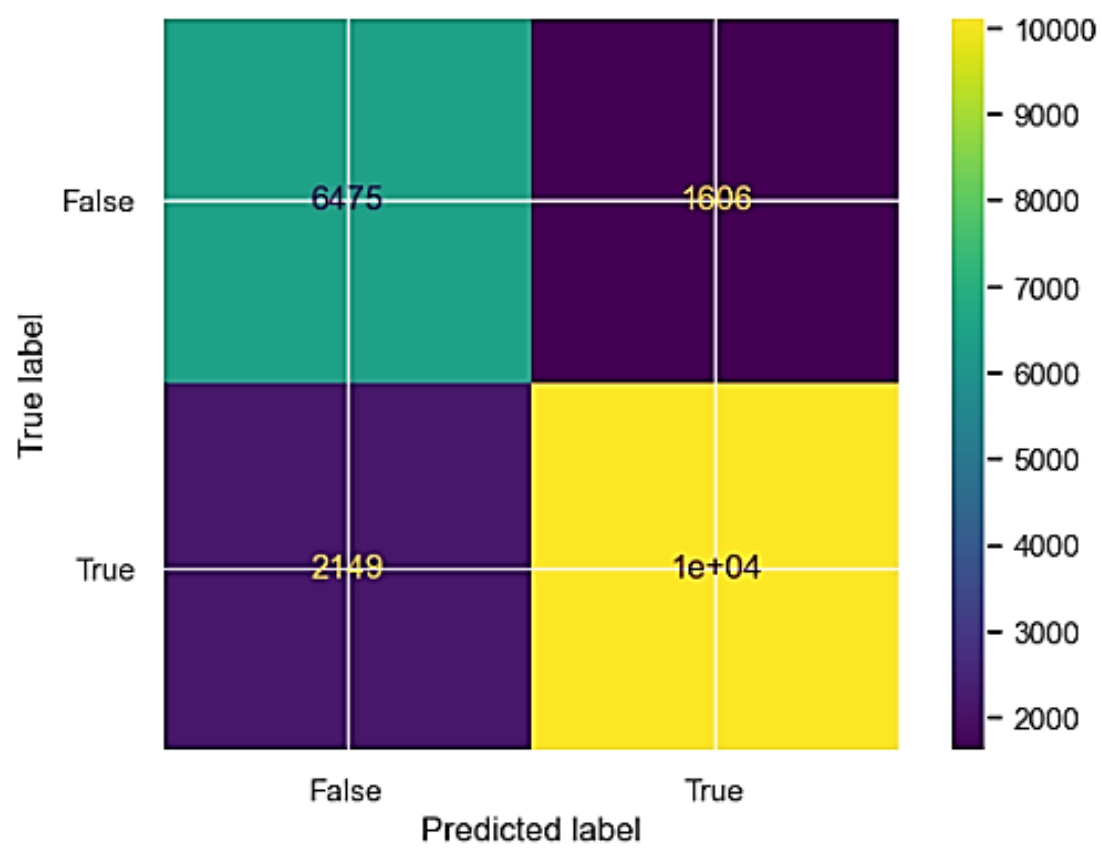

Figure 7. Confusion matrix of the RFC based on  transformed MORGAN2 dataset

Moreover, when the RFC was applied to the dataset transformed using RDKit SMILES, it achieved an Accuracy of 89.2%, a Precision of 91.8%, a Recall of 90.7%, an F1-score of 90.1% and an ROC of 89%. Tracking for the point where the deviation between train and test accuracy was higher than 5% point max_depth=15, where the train Accuracy was 90.1% and the test Accuracy was 84.7%. So, the choice was max_depth=14 with train Accuracy 89.3% and test Accuracy 84.5%. This process, which guided the model's optimisation, is visually shown in Figure 8.

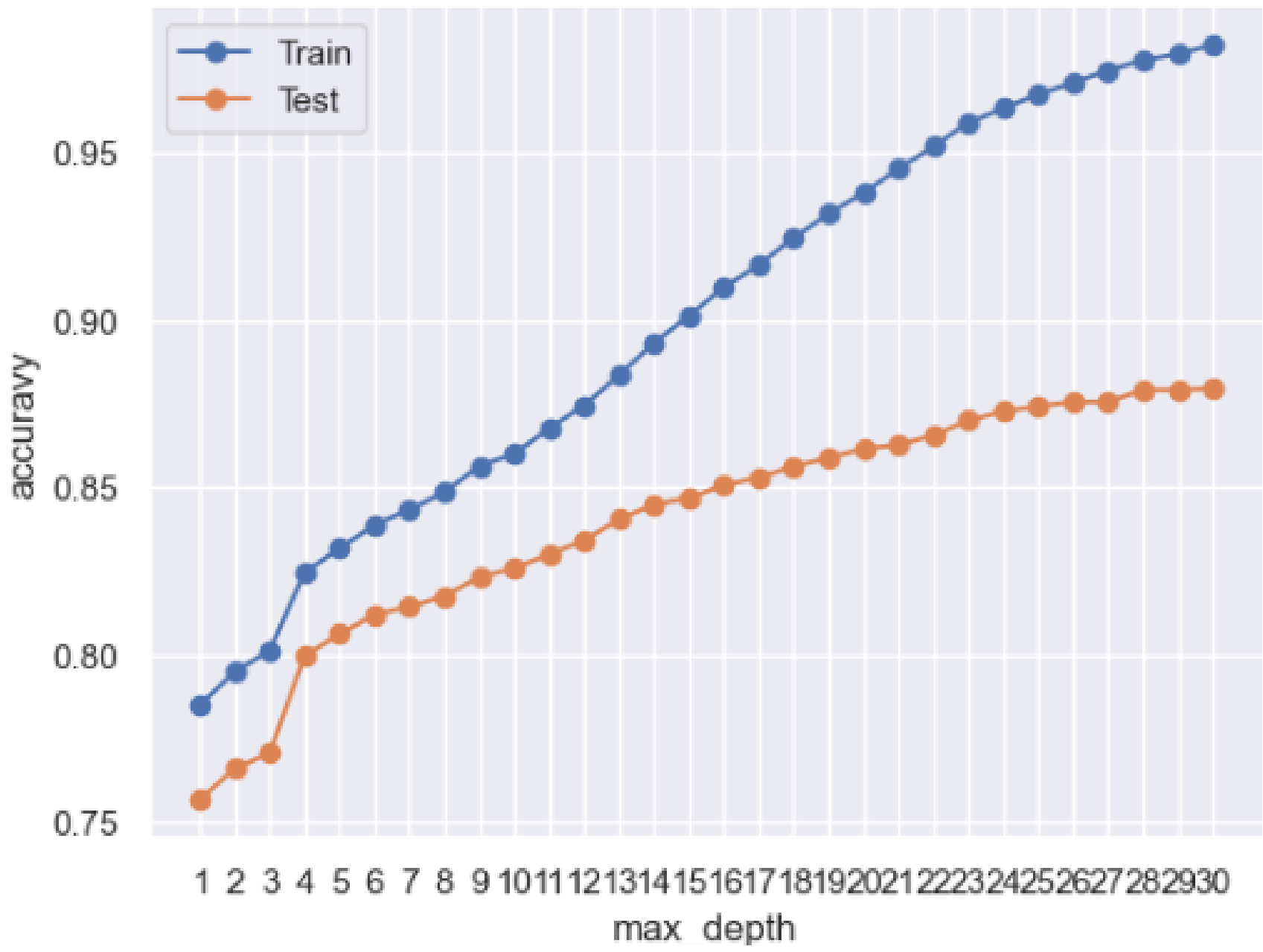

Figure 8.
Overfitting analysis of the RFC based on dataset with transformed by RDKit SMILES: prediction accuracy vs maximum depth of the decision tree. The blue line is the training accuracy. The orange line is the test accuracy. The deviation between the testing and training accuracy higher than 5% was considered as an indication for early stopping.

The RFC configured with a maximum depth of 14 yielded a highly stable and strong final performance on the test set. It achieved an Accuracy of 84.1%, a Precision of 91.1%, a Recall of 81.6%, an F1-score of 86.1% and an ROC of 84.7%. The model's consistency is confirmed by the fact that these single-run metrics are all within a narrow range (0.17% to 0.25%) of their respective CV averages. This close agreement, further supported by the CV's low standard deviations (all below ±0.25%), indicates the model exhibits low variance and reliable generalisation without signs of overfitting. The performance profile highlights a preference for high Precision over Recall, making its positive predictions highly trustworthy. Both the CV and final results highlight a strong preference for Precision (~91%) over Recall (~ 82%), indicating the model is highly trustworthy when predicting a positive outcome. However, the ROC shows a notable drop from the CV average of 92.23% ± 0.17% to 84.7% on the test set, which suggests that while the model's performance is excellent at its specific classification threshold, its overall discriminative ability across all thresholds is significantly reduced on the final test data.

Table 5. Five-fold cross-validation of RFC based on MORGAN2 transformed features with StratifiedKFold

| **Metrics** | **Fold** | | | | | **Average across the folds** |
|---|---|---|---|---|---|---|
| | **1** | **2** | **3** | **4** | **5** | |
| **Accuracy [%]** | 83.46 | 84.16 | 83.94 | 83.89 | 83.70 | 83.83±0.23 |
| **Precision [%]** | 90.38 | 90.97 | 91.00 | 90.80 | 90.84 | 90.80±0.22 |
| **Recall [%]** | 81.24 | 81.87 | 81.43 | 81.57 | 81.17 | 81.45±0.25 |
| **F1-Score [%]** | 85.57 | 86.18 | 85.95 | 85.94 | 85.73 | 85.87±0.21 |
| **ROC AUC [%]** | 92.27 | 92.46 | 92.29 | 92.19 | 91.94 | 92.23±0.17 |

| | | | | | | | |
|---|---|---|---|---|---|---|---|
| **Confusion Matrix** | **TN** | 5615 | 5666 | 5673 | 5651 | 5659 | 5652.8±20.26 |
| | **FP** | 850 | 799 | 792 | 813 | 805 | 811.8±22.71 |
| | **FN** | 1845 | 1783 | 1826 | 1812 | 1852 | 1823.6±27.63 |
| | **TP** | 7988 | 8050 | 8007 | 8021 | 7981 | 8009.4±27.63 |

The classification results indicate that the RFC based on a dataset on RDKit SMILES was performing solidly and reliably. The model exhibits a strength in identifying positive cases, correctly predicting 10,086 TP (Figure 9). The high resulting Precision is strongly supported by the low number of incorrect positive predictions, with only 1,052 FP. However, the primary area for improvement is the substantial number of 2,205 missed positive cases FN, which impacts the model's Recall. The overall performance is quantified by the high F1-score of 0.8610 and an MCC of 0.6785, confirming a strong, positive correlation between the predicted and true classifications and demonstrating good performance across both class predictions.

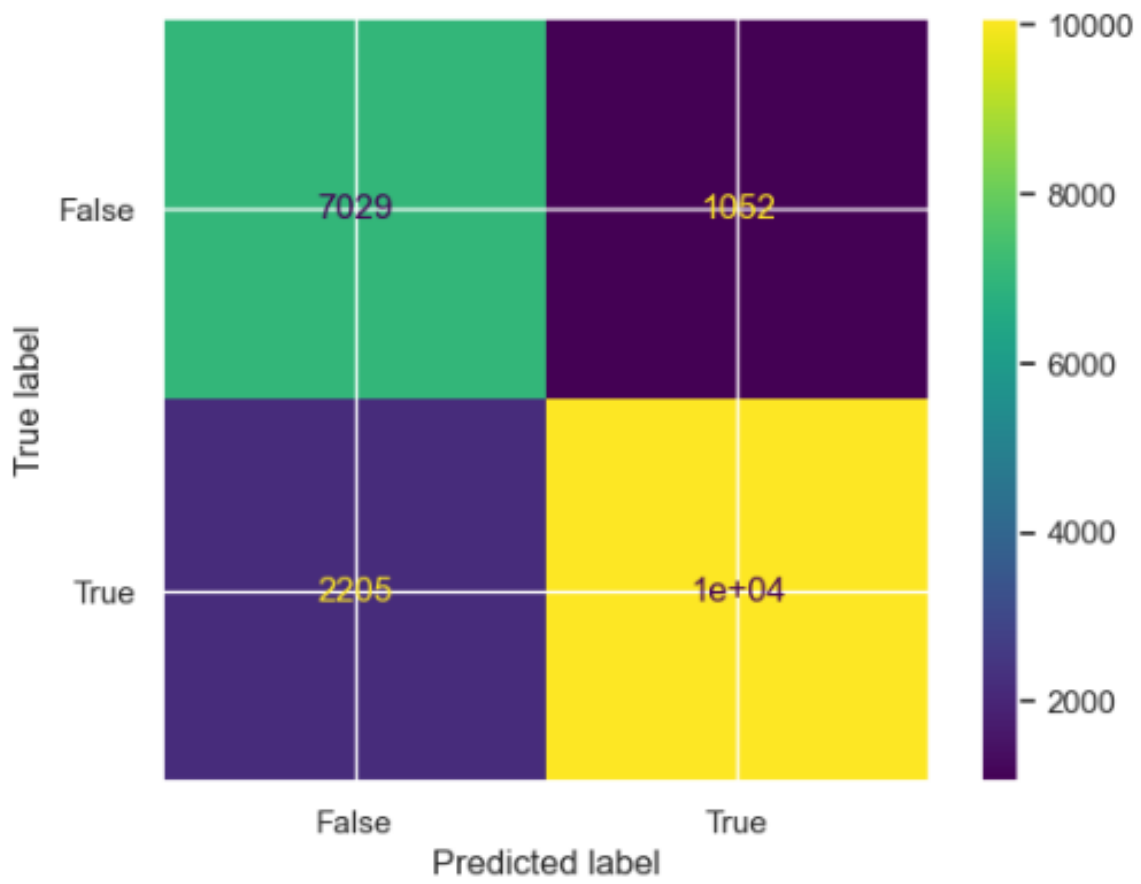

Figure 9. Confusion matrix of the RFC based on dataset with transformed by RDKit SMILES

The comparative performance, as measured by the MCC, shows a clear hierarchy among the RFC models based on their feature encoding methods. The RFC using IUPAC tokenised data performed the weakest with an MCC of 0.4466, indicating a moderate correlation barely better than random. Performance significantly improved when using the MORGAN2 features, yielding an MCC of 0.6202, suggesting a strong positive correlation and a more reliable model. The best performance, however, was achieved by the RFC trained on RDKit-converted SMILES features, which reached the highest MCC of 0.6785, confirming that this encoding method provided the most balanced and strongest prediction quality across both positive and negative classes for the classification task.

Overall, while the ML models based on SMILES inherently demonstrated superior performance metrics, the model utilizing IUPAC-tokenised data offers a distinct and significant advantage: a relatively straightforward path to providing human-readable insights for drug discovery. This direct interpretability allows medicinal chemists to immediately identify potential functional groups and structural fragments that drive activity, enabling targeted design efforts. Conversely, extracting comparable chemical intelligence from an SMILES-based ML model requires complicated post-hoc analysis and additional calculations to translate the abstract molecular strings into meaningful chemical features, making the IUPAC approach a far more practical and time-efficient tool for human decision-making in the initial phases of drug development

*Identification of the relevant functional groups for the inhibitory action*

The feature importance analysis indicates which functional groups could be the most relevant for a specific chemical interaction. In Figure 10 below, a list of 24 functional groups with the highest relevance with respect to the inhibition of TDP1 is shown. It should be noted that the computations were based on the mutual influence between all features. Here, the first 24 functional groups out of the list of 5,963 groups are shown.

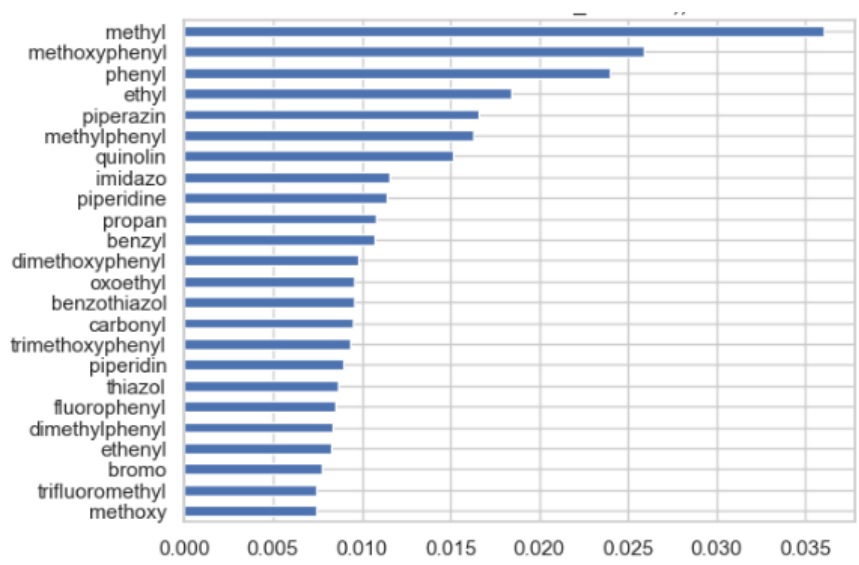

a)

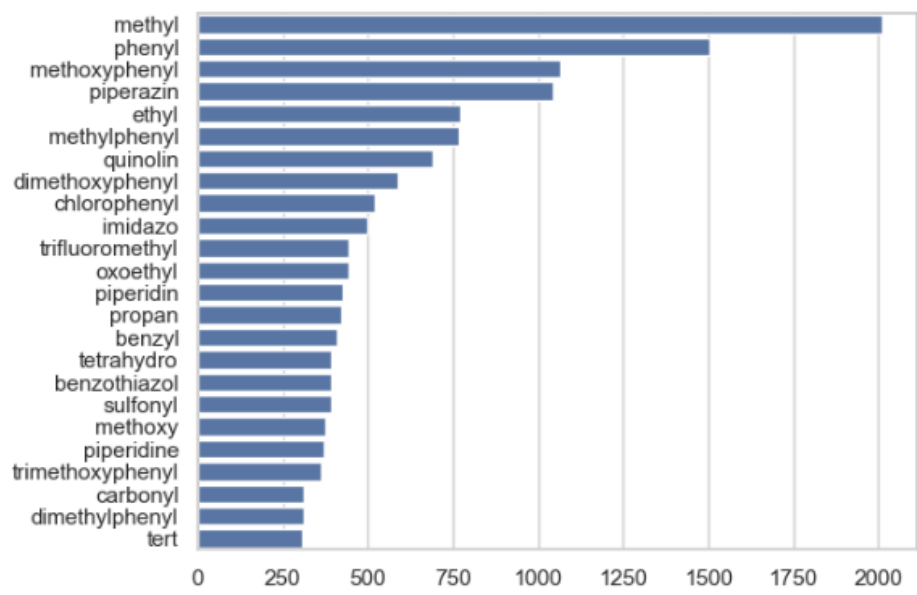

b)

Figure 10. Feature (functional group) importance ranking for the prediction of TDP1 inhibitors.
a) Scikit Learn Feature Importance algorithm for the Random Forest Classifier. b) Chi2 algorithm.

Feature importance in a binary classifier quantifies each feature's contribution to the model's overall predictive power, but it doesn't specify whether a feature characterises the active or inactive class. While machine learning algorithms identify functional groups correlated with inhibition, this correlation doesn't guarantee the group is actively contributing to the effect; some groups may frequently appear in non-inhibitors. Therefore, to prioritise functional groups most likely to contribute to true inhibitors, we calculated the ratio (or relative proportion) of active versus inactive compounds containing that specific group. This ratio serves as a refined metric to focus on structurally enriched features.

The combined analysis of functional groups provides a nuanced view of the structural features governing activity, based on two ranking methods (Table 6). The imidazo group stands out as the most potent statistical predictor, with 89% of compounds containing it being active and an Active/Inactive Ratio of 8.33, though it only ranks 9th in Feature Importance. In contrast, the general methyl group is the model's top feature (Rank 1), despite a lower activity concentration (69%), likely due to its high overall count (20,205 active cases), making it crucial for the model's predictive power. The strong statistical enrichment of groups like ethenyl (87%) and quinolin (86%) confirms they are highly reliable activity markers, while groups such as carbohydrazide (22% active, Ratio 0.27) are powerful indicators of inactivity. This difference in rankings highlights that the Relative Proportion identifies the most enriched groups, while the Feature Importance identifies the most useful features for the specific ML classifier. A check was performed on a random sample of five compounds containing an *imidazo*, and all were determined not to be PAINs (Table 7). A comprehensive study of all samples containing an *imidazo*, as well as the similarities and differences of these structures, is a subject of further research.

Table 6. Relevance of functional groups regarding the prediction of TDP1 inhibitors, by using the Feature Importance algorithm (column 1), and reordered (in col. 2) on the basis of the value for the ratio of the number of Active inhibitor substances (col. 4) and Inactive (col.5), which is give in col.6.

| Position according to Feature Importance algorithm | **Position according to relative proportion of the active cases** | Functional group/fragment | Number of substances that are active inhibitors and contain this group (Active cases) | Number of substances that are not active inhibitors and contain this group (Inactive cases) | Percentage of Active cases | **Ratio of Active and Inactive cases for the functional group** |
|---|---|---|---|---|---|---|
| 9 | **1** | *imidazo* | 1050 | 126 | 89% | **8.33** |
| 21 | **2** | *ethenyl* | 591 | 85 | 87% | **6.95** |
| 20 | **3** | *trimethoxyphenyl* | 731 | 119 | 86% | **6.14** |
| 6 | **4** | *quinolin* | 1785 | 299 | 86% | **5.97** |
| 3 | **5** | *piperazin(e)** | 4211 | 870 | 83% | **4.84** |
| 24 | **6** | *tetrahydro* | 1259 | 302 | 81% | **4.17** |
| 15 | **7** | *benzothiazol* | 1375 | 368 | 79% | **3.74** |
| 12 | **8** | *sulfonyl* | 1512 | 427 | 78% | **3.54** |
| 13 | **9** | *piperidine* | 1839 | 547 | 77% | **3.36** |
| 10 | **10** | *dimethoxyphenyl* | 2647 | 806 | 77% | **3.28** |
| 25 | **11** | *methanone* | 1831 | 596 | 75% | **3.07** |
| 2 | **12** | *phenyl* | 8524 | 2941 | 74% | **2.90** |
| 19 | **13** | *piperidin* | 2457 | 851 | 74% | **2.89** |
| 17 | **14** | *trifluoromethyl* | 2406 | 865 | 74% | **2.78** |
| 14 | **15** | *benzyl* | 2458 | 892 | 73% | **2.76** |
| 4 | **16** | *methoxyphenyl* | 6966 | 2559 | 73% | **2.72** |
| 8 | **17** | *methylphenyl* | 5961 | 2383 | 71% | **2.50** |
| 11 | **18** | *chlorophenyl* | 4749 | 2004 | 70% | **2.37** |
| 7 | **19** | *ethyl* | 8252 | 3752 | 69% | **2.20** |
| 1 | **20** | *methyl* | 20205 | 9283 | 69% | **2.18** |
| 16 | **21** | *carboxamide* | 9495 | 4910 | 66% | **1.93** |
| 5 | **22** | *oxoethyl* | 2420 | 3059 | 44% | **0.79** |
| 22 | **23** | *acetate* | 655 | 982 | 40% | **0.67** |
| 18 | **24** | *carbohydrazide* | 126 | 460 | 22% | **0.27** |

As noted earlier, the IUPAC names were subjected to direct tokenisation, with the resultant components being utilized without modification. Specifically, this process did not generate new string elements or introduce any splitting of the strings beyond the divisions already established by the IUPAC nomenclature rules. To ensure data consistency and accurate representation, the mean value should be calculated for any functional group or molecular fragment that is observed with different spellings in the tokenized results. For example, instances like 'piperazin' and 'piperazine' should be collated and their mean value determined. For example, '*piperazin*' and 'piperazine' have the relative proportion values of 5.06 and 4.29, respectively. So, the relative proportion of the active and inactive cases for '*piperazine*' is calculated to be 4.68. That process automation is an object for further development.

Table 7. Results of checking five random samples containing *imidazo* if they are PAINs

| CID | SID | IUPAC | PAIN |
|---|---|---|---|
| 3245566 | 4251947 | 5-ethyl-N-(2-imidazo[1,2-a]pyridin-2-ylethyl)thiophene-2-sulfonamide | No |
| 20908515 | 49736571 | 2-(1H-imidazo[4,5-b]pyridin-2-ylsulfanyl)-N-(2-phenylethyl)butanamide | No |
| 844907 | 7974454 | 2-(1H-imidazo[4,5-b]pyridin-2-ylsulfanyl)-N-(2-methoxyphenyl)acetamide | No |
| 7066893 | 24269226 | 2-methoxy-4-[5-(3-methylanilino)imidazo[2,1-b][1,3]thiazol-6-yl]phenol | No |
| 16018392 | 24396339 | N-(4-methoxyphenyl)-3-(4-methylphenyl)-6,7,8,9-tetrahydro-5H-imidazo[1,5-a]azepine-1-carboxamide | No |

The Fisher's Exact Test analysis demonstrates that the presence of every functional group listed has a highly significant statistical association with a substance's activity status, as all P-values are extremely small ($P < 0.0021$). The Odds Ratio (OR) indicates the strength and direction of this relationship: the imidazo group (OR = 5.555) shows the strongest positive association, making a compound over five times more likely to be Active if this group is present. Other strong positive predictors include ethenyl (OR = 4.605) and trimethoxyphenyl (OR = 4.074). Conversely, several groups are strongly associated with inactivity (OR < 1.0); the carbohydrazide group (OR = 0.178) has the strongest negative association, making a substance with this feature about 5.6 times more likely to be Inactive, followed by acetate (OR = 0.432) and oxoethyl (OR = 0.500). This analysis clearly identifies key structural features that either strongly promote or strongly inhibit activity.

Table 8. The Fisher's Exact Test analysis was applied to the functional groups/fragments that were identified as having the highest feature importance for the ML models

| Functional Group | Active (a) | Inactive (b) | Odds Ratio | P-value | Significance |
|---|---|---|---|---|---|
| carboxamide | 9495 | 4910 | 57.78 | 0.00e+00 | *** Highly Significant(P < 0.0021) |
| phenyl | 8524 | 2941 | 33.50 | 0.00e+00 | *** Highly Significant (P < 0.0021) |
| ethyl | 8252 | 3752 | 20.44 | 0.00e+00 | *** Highly Significant (P < 0.0021) |
| imidazo | 1050 | 126 | 18.50 | 0.00e+00 | *** Highly Significant (P < 0.0021) |
| piperazin(e)* | 4211 | 870 | 15.99 | 0.00e+00 | *** Highly Significant (P < 0.0021) |
| methoxyphenyl | 6966 | 2559 | 15.64 | 0.00e+00 | *** Highly Significant (P < 0.0021) |
| ethenyl | 591 | 85 | 14.71 | 2.04e-192 | *** Highly Significant (P < 0.0021) |
| quinolin | 1785 | 299 | 14.31 | 0.00e+00 | *** Highly Significant (P < 0.0021) |
| trimethoxyphenyl | 731 | 119 | 13.17 | 4.28e-229 | *** Highly Significant (P < 0.0021) |
| methylphenyl | 5961 | 2383 | 10.91 | 0.00e+00 | *** Highly Significant (P < 0.0021) |
| tetrahydro | 1259 | 302 | 9.39 | 0.00e+00 | *** Highly Significant (P < 0.0021) |
| dimethoxyphenyl | 2647 | 806 | 8.57 | 0.00e+00 | *** Highly Significant (P < 0.0021) |
| benzothiazol | 1375 | 368 | 8.50 | 0.00e+00 | *** Highly Significant (P < 0.0021) |
| sulfonyl | 1512 | 427 | 8.16 | 0.00e+00 | *** Highly Significant (P < 0.0021) |
| chlorophenyl | 4749 | 2004 | 8.12 | 0.00e+00 | *** Highly Significant (P < 0.0021) |
| piperidine | 1839 | 547 | 8.01 | 0.00e+00 | *** Highly Significant (P < 0.0021) |
| piperidin | 2457 | 851 | 7.33 | 0.00e+00 | *** Highly Significant (P < 0.0021) |
| methanone | 1831 | 596 | 7.30 | 0.00e+00 | *** Highly Significant (P < 0.0021) |
| trifluoromethyl | 2406 | 865 | 7.01 | 0.00e+00 | *** Highly Significant (P < 0.0021) |
| benzyl | 2458 | 892 | 6.98 | 0.00e+00 | *** Highly Significant (P < 0.0021) |
| oxoethyl | 2420 | 3059 | 1.77 | 1.72e-76 | *** Highly Significant (P < 0.0021) |
| acetate | 655 | 982 | 1.36 | 6.40e-09 | *** Highly Significant (P < 0.0021) |
| carbohydrazide | 126 | 460 | 0.54 | 2.61e-10 | *** Highly Significant (P < 0.0021) |

A check was performed on a random sample of five compounds containing a *carboxamide*, and all were determined not to be PAINs (Table 9). A comprehensive study of all samples containing a *carboxamide*, as well as the similarities and differences of these structures, is a subject of further research.

Table 9 Results of checking five random samples containing *carboxamide* if they are PAINs

| CID | SID | IUPAC | PAIN |
|---|---|---|---|
| 15993269 | 49667698 | 5-(1,3-benzodioxol-5-yl)-N-(2-methylpropyl)-1,2-oxazole-3-carboxamide | No |
| 16008588 | 24384816 | 1-(2-chlorophenyl)-N-(3,5-dimethoxyphenyl)-3,6-dimethylpyrazolo[3,4-b]pyridine-4-carboxamide | No |
| 16014225 | 24391772 | 2-(4-ethylphenyl)-5-(hydroxymethyl)-N-(thiophen-2-ylmethyl)triazole-4-carboxamide | No |
| 4086767 | 24415205 | 4-benzyl-N-[2-(4-chlorophenyl)ethyl]-3-oxo-1,4-benzothiazine-6-carboxamide | No |
| 46904149 | 99359587 | 1-benzyl-6-methyl-2-oxo-3-[2-oxo-2-(4-phenylbutylamino)ethyl]-N,N-di(propan-2-yl)-3,4-dihydropyridine-5-carboxamide | No |

Boruta feature selection algorithm failed to identify any statistically relevant features whose importance was significantly higher than random noise (the "shadow features"). In essence, the algorithm suggests that none of the initial features contain a detectable signal that is stronger than pure chance for the given prediction task. Because the subsequent stage relies on a confirmed subset of features to build the RFC, the process cannot proceed as intended, as the feature set is considered uninformative.

### Results regarding the second methodology

There were other ways of ranking the functional groups. For example, the ranking can be based on their participation in only one type of cases (active or inactive):

(i) For the functional groups that participate only in active (i.e. it is a TDP1 inhibitor) small biomolecule content and in no inactive compounds, the leading functional group was *oxonaphthalen* with 25 active cases, followed by *methylsulfonylpyrimidine* with 22 and *tetrahydroindol* with 20 active cases. The entire ranking list of 2,178 functional groups participating in the content of the active small biomolecule is available on GitHub [43]. A check was performed on a random sample of five compounds containing the *oxonaphthalen* and four out of five were flagged as PAINs (Table 10). A comprehensive study of all samples containing an *oxonaphthalen*, as well as the similarities and differences of these structures, is a subject of further research.

Table 10 Results of checking five random samples containing *oxonaphthalen* if they are PAINs

| CID | SID | IUPAC | PAIN |
|---|---|---|---|
| 752424 | 24809810 | 1,5-dimethyl-4-[(4-oxonaphthalen-1-ylidene)amino]-2-phenylpyrazol-3-one | PAIN |
| 6032979 | 17511248 | (NZ)-N-[3-(4-methylanilino)-4-oxonaphthalen-1-ylidene]thiophene-2-sulfonamide | PAIN |
| 5676317 | 17517158 | (NZ)-N-(3-anilino-4-oxonaphthalen-1-ylidene)thiophene-2-sulfonamide | PAIN |
| 4441046 | 14742503 | N-(1-dibutoxyphosphoryl-4-oxonaphthalen-1-yl)benzenesulfonamide | No |
| 5105556 | 87347760 | N-[3-bromo-1-di(propan-2-yloxy)phosphoryl-4-oxonaphthalen-1-yl]benzenesulfonamide | PAIN |

(ii) For the functional groups that participate only in inactive (i.e. it is not a TDP1 inhibitor) small biomolecule composition and in none of the active, on top of this list was *ylbutanediamide* with 104 inactive cases, followed by *oxopiperazin* with 100 and *tetrazabicyclo* with 99. The entire ranking list of 6,243 functional groups that participated only in the content of the inactive small biomolecule is available on GitHub [44]. A check was performed on a random sample of five compounds containing a *ylbutanediamide*, and all were determined not to be PAINs (Table 11). A comprehensive study of all samples containing a *ylbutanediamide*, as well as the similarities and differences of these structures, is a subject of further research.

Table 11 Results of checking five random samples containing *ylbutanediamide* if they are PAINs

| CID | SID | IUPAC | PAIN |
|---|---|---|---|
| 3205145 | 14721697 | N'-(2,3-dihydro-1,4-benzodioxin-6-yl)-N'-[2-(3-methylbutylamino)-2-oxoethyl]-N-pyridin-2-ylbutanediamide | No |
| 654170 | 26668007 | N'-[2-(2-methoxyethylamino)-1-(4-methoxyphenyl)-2-oxoethyl]-N'-(oxolan-2-ylmethyl)-N-pyridin-2-ylbutanediamide | No |
| 651822 | 26668581 | N'-[1-(4-fluorophenyl)-2-(2-methylbutan-2-ylamino)-2-oxoethyl]-N'-(furan-2-ylmethyl)-N-pyridin-2-ylbutanediamide | No |
| 3205015 | 49725897 | N'-[2-(tert-butylamino)-1-(4-methoxyphenyl)-2-oxoethyl]-N'-cyclohexyl-N-pyridin-2-ylbutanediamide | No |
| 3204777 | 49726932 | N'-(4-methoxyphenyl)-N'-[2-oxo-2-(2-phenylethylamino)ethyl]-N-pyridin-2-ylbutanediamide | No |

The rationale for not excluding PAINs (Pan-Assay Interference Compounds) from the initial dataset was explained in the Methodology section. However, out of curiosity, these PAINs were subsequently flagged using the PubChem AID 686978 bioassay data [11]. This flagging process resulted in 21,761 samples being identified as PAINs, leaving 388,803 compounds that were revealed as non-PAINs. This represents a 6.46% decrease in the total samples, which is unevenly spread across the classes, as shown in Table 12. Although no dramatic decrease was observed, the resulting PAIN-free dataset was considered of sufficient interest for further investigation [45 , 46].

Table 12. Result of application of a PAIN filter on the PubChem AID 686978 bioassay`s dataset

| **Samples** | **Before the PAIN filter** | **After the PAIN filter** | **Decrease** |
|---|---|---|---|
| Inactive | 236,226 | 227,158 | 3.84% |
| Inconclusive | 112,867 | 105,950 | 6.13% |
| Active | 61,471 | 55,695 | 9.40% |

As an option, given the correlation between the chemical shifts of a biomolecule provided by the 13C NMR spectroscopy and its functionality [47, 48], it was hypothesised that when a tested compound contains one of the extracted functional groups/fragments and its 13C NMR spectroscopy data resembles of the 13C NMR spectroscopy data of the source compound of this functional group/fragments, there is a high probability that the tested compound is a TDP1 inhibitor. One of the tools that can provide such a comparison of the NMR spectroscopy data is the ACD/Labs [49].

### *Results regarding the third methodology*

The CID_SID ML model that was developed beyond the main study to aid drug discovery researchers interested in TDP1 inhibition, achieved with the XGBC Accuracy of 86.1%, Precision of 93.3%, Recall of 77.8%, F1-score of 84.9%, ROC of 86.1%, followed by GBC with Accuracy of 85.2%, Precision of 94.2%, Recall of 75.0%, F1-score of 83.5%, ROC of 85.2% (Table 13). The results were achieved by training the ML model with 100,942 samples and tested with 22,000 samples [50].

Table 13. ML metric regarding ML models based on CID and SID model predicting TDP1 inhibition

| **Algorithm** | **Accuracy** | **Precision** | **Recall** | **F1-score** | **ROC** |
|---|---|---|---|---|---|
| XGBoost | 0.861 | 0.933 | 0.778 | 0.849 | 0.861 |
| GradientBoost | 0.852 | 0.942 | 0.750 | 0.835 | 0.852 |
| RandomForest | 0.846 | 0.855 | 0.835 | 0.845 | 0.846 |
| K-nearest | 0.832 | 0.844 | 0.814 | 0.829 | 0.832 |
| Decision | 0.800 | 0.770 | 0.856 | 0.810 | 0.800 |
| SVM | 0.792 | 0.912 | 0.645 | 0.756 | 0.792 |

A statistical significance test was conducted to compare the two machine learning classifiers, the GBC (Model A) and XGBC (Model B), and it was concluded that Model B is the statistically superior performer. While both models achieved high accuracy (Model A: 0.8519, Model B: 0.8611), the small difference in favour of XGBC was confirmed as significant by McNemar's Test, which yielded a P-value of 0.0000. This result, far below the 0.05 significance level, indicates the performance difference is not due to random chance. Further analysis of the disagreement counts supports this finding: Model B correctly classified 583 samples that Model A missed, while Model A only correctly classified 379 samples that Model B missed, showing a clear and statistically validated advantage for the XGBoost implementation.

The tracing of accuracy deviation during hyperparameter tuning revealed the point where model complexity began to hinder generalisation (Figure 11). The highest deviation between training and testing accuracy was observed at max_depth = 22, where the model achieved a training accuracy of 89.6% and a test accuracy of 84.4%, indicating an undesirable degree of overfitting (a 5.2% gap). Crucially, the best test performance was achieved at a simpler setting, max_depth=7, which yielded a test Accuracy of 86.2%. At this optimal depth, the training Accuracy was 86.2%, resulting in a much smaller and healthier deviation of 2.9%. This confirms that max_depth=7 represents the optimal balance point, providing the highest generalisation capability before the model started to memorise noise instead of learning general patterns.

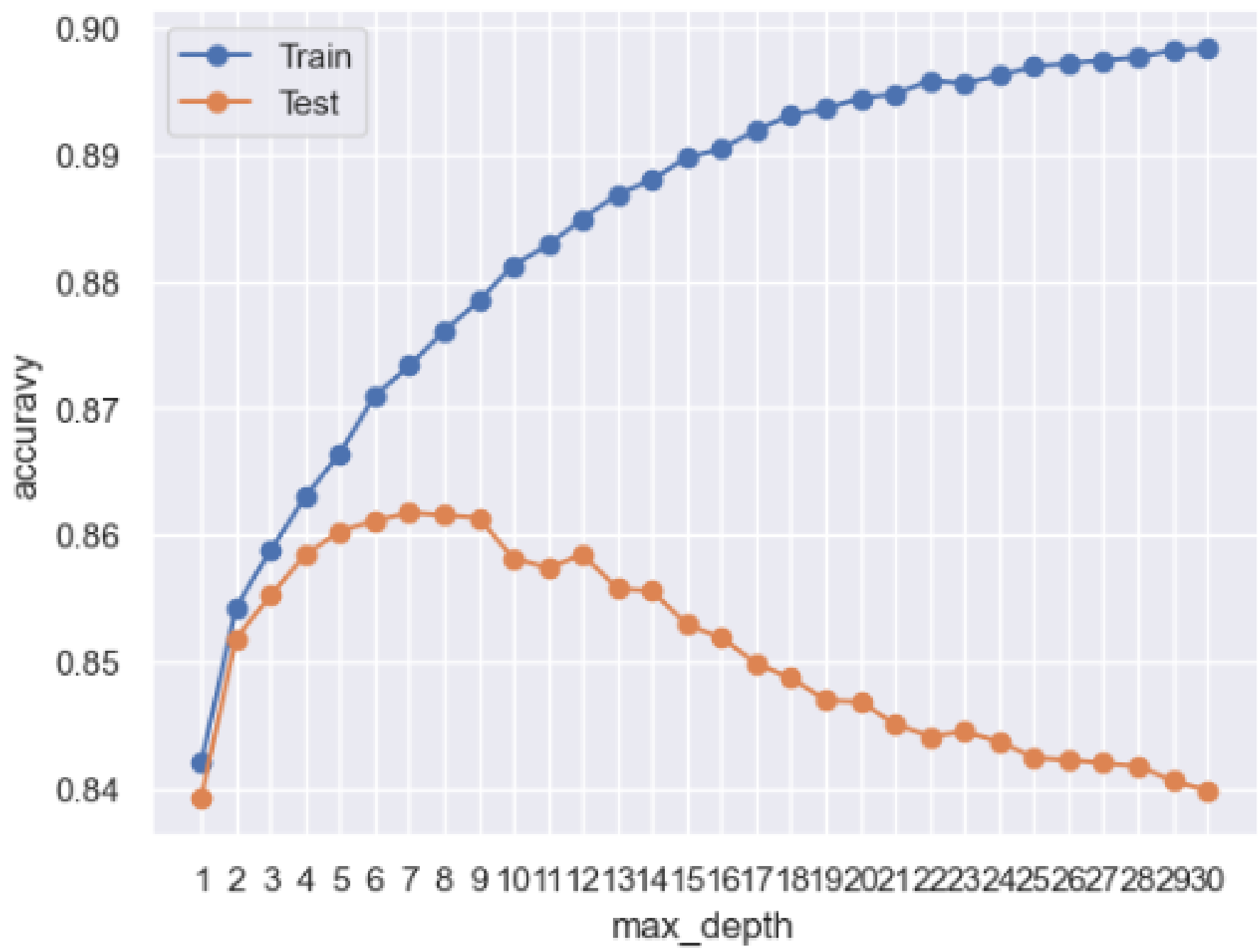

Figure 11. Scrutinizing for overfitting of the CID_SID XGBC ML model that predicts the TDP1 inhibitors. The blue line is the train accuracy. The orange line is the test accuracy. The deviation between the test and train accuracy higher than 5% is an indication for overfitting.

As was noted above, the XGBC model's performance, when evaluated as a single run, showed an Accuracy of 86.1%, a Precision of 93.3%, a Recall of 77.8%, an F1-score of 84.9% and an ROC of 86.1%. However, a more robust assessment using five-fold CV revealed a slightly lower but more reliable average Accuracy of 84.01% (±0.17%) and a marginally lower Precision of 95.42% (± 0.25%). Notably, the CV summary indicated a slightly higher Recall of 78.45% (± 0.14%) and a better average F1-score of 86.11% (±0.14%), suggesting the model is generally slightly better balanced and more effective across different data partitions. Most strikingly, the CV average ROC saw a significant increase to 91.84% (± 0.22%), indicating that while the single run was a good estimate, the CV results provide a more optimistic and statistically stable measure of the model's discriminative power across various thresholds. The results are shown in detail in Table 14.

Table 14 Five-fold cross-validation results for the CID_SID machine learning models with XGB algorithm

| Metrics | Fold | Average |
|---|---|---|

| | | 1 | 2 | 3 | 4 | 5 | across the folds |
|---|---|---|---|---|---|---|---|
| **Accuracy [%]** | | 83.77 | 83.96 | 83.99 | 84.01 | 84.29 | 84.01±0.17 |
| **Precision [%]** | | 95.20 | 95.27 | 95.28 | 95.33 | 95.91 | 95.42±0.25 |
| **Recall [%]** | | 78.18 | 78.50 | 78.56 | 78.54 | 78.49 | 78.45±0.14 |
| **F1-Score [%]** | | 85.89 | 86.08 | 86.12 | 86.13 | 86.33 | 86.11±0.14 |
| **ROC AUC [%]** | | 91.75 | 91.86 | 91.52 | 91.86 | 92.21 | 91.84±0.22 |
| **Confusion Matrix** | **TN** | 5491 | 5488 | 5488 | 5493 | 5543 | 5500±21.28 |
| | **FP** | 390 | 393 | 393 | 388 | 338 | 380.4±21.28 |
| | **FN** | 2203 | 2170 | 2164 | 2166 | 2171 | 2174±14.33 |
| | **TP** | 7892 | 7924 | 7930 | 7928 | 7923 | 7919.4±13.94 |

From one hundred trials, the optimisation framework Optuna successfully selected the best set of hyperparameters for the XGBC. The optimal configuration included a maximum tree depth of 10 ('max_depth': 10), a reduced learning rate of approximately 0.017, 463 estimators ('n_estimators': 463), and specific regularization and sampling values ('gamma': 0.389, 'reg_lambda': 0.316, 'min_child_weight': 4). The Optuna hyperparameter optimization run resulted in a model with a slightly lower Accuracy of 0.856 compared to the model using default hyperparameters, which achieved an Accuracy of 0.862. This suggests that the optimisation process either failed to find a better configuration than the default one or, perhaps more concerningly, settled on a less effective set of hyperparameters for maximising this specific metric. The difference, while small (0.006), indicates that the default settings were, in this instance, superior for classification accuracy.

The 95% CI, derived from 1000 bootstraps, provides crucial context: it indicates that the true performance of the model is highly likely to fall within the tight range of 0.8574 to 0.8667. This narrow range, with a width of approximately 0.0093, confirms the stability and robustness of the model's performance.. Its Precision is particularly impressive at 0.9299 (95% CI: 0.9247 to 0.9353), suggesting that when the model predicts a positive outcome, it is correct nearly 93% of the time, resulting in a very low FP rate. The Recall metric, however, is comparatively lower at 0.7825 (95% CI: 0.7753 to 0.7903), meaning approximately 78% of all true positive cases were correctly identified, leaving a notable portion (about 21.7%) as FN. Overall, the XGBC model is highly reliable in its positive predictions, and the narrow confidence intervals across all metrics confirm the stability and robustness of its performance across bootstrapped samples.

The performance of the XGBC model demonstrates a substantial improvement over the baseline RFC model, confirming its superior capability for classifying TDP1 inhibitors. The narrow 95% CI, calculated from 1000 bootstraps on a test set of 22,000 samples, attests to the high reliability and stability of these metrics. Specifically, the Accuracy CI of 0.8577 to 0.8665 indicates the model is highly likely to be correct about 86% of the time. Crucially, the Precision CI of 0.9280 to 0.9385 shows that when XGBC predicts a compound is a positive inhibitor, it is correct over 93% of the time, resulting in a significantly reduced false positive rate (~ 7\%) that is highly desirable for minimising the screening of inactive compounds in expensive wet-lab experiments. The Recall CI of 0.7725 to 0.7870 means the model successfully identifies nearly 78% of the actual inhibitors, a robust improvement that, while not perfect, is strong enough to capture a large fraction of active compounds for further drug discovery efforts.

The XGBC model based on CIDs and SIDs data demonstrates strong and balanced predictive power, highlighted by an excellent MCC of 0.7328. The raw counts, 10,351 TN and 8,608 TP, confirm the model's high overall accuracy in correctly identifying both classes. A low count of 649 FP suggests high Precision (the model rarely raises a false alarm), while the 2,392 FN indicate that the main challenge lies in improving Recall (the rate at which it correctly captures all positive cases). Overall, the high MCC value, which is robust against class imbalance, validates the model as a highly effective and reliable solution for the classification task (Figure 12).

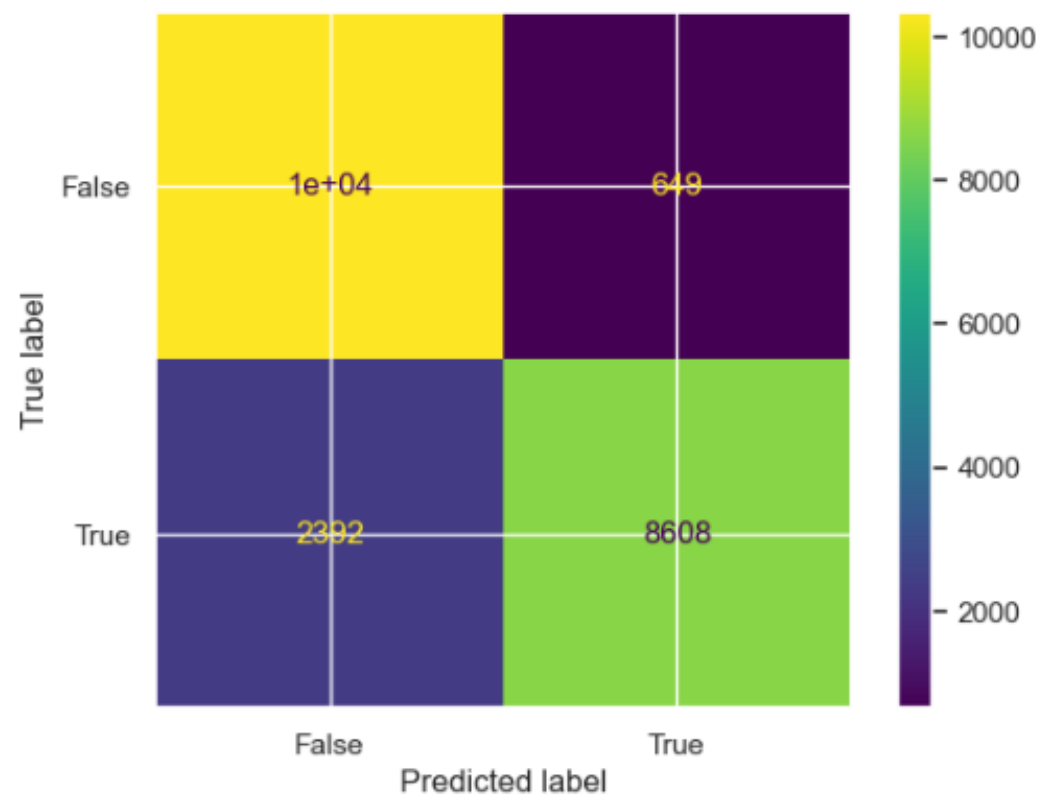

Figure 12. The CID_SID XGBC ML model confusion matrix

The classification report (Table 15) demonstrates that the model achieved strong, balanced performance with an overall Accuracy of 0.86 on a perfectly balanced dataset (11,000 samples per class). The model exhibits a classic Precision-Recall trade-off across the two classes: it is highly effective at finding positive cases, indicated by the exceptional Recall of 0.94 for the Active class (Target 1). This high recall, however, comes at the cost of its precision, resulting in a moderate Precision of 0.81 for that same class, meaning it has a higher rate of "false alarms" when predicting active instances. Conversely, the model is very conservative and highly trustworthy when predicting the negative class, boasting Precision of 0.93 for the Inactive class (Target 0), though its ability to find all truly inactive cases is lower, with a Recall of 0.78. Overall, the macro and weighted average F1-scores of 0.86 confirm the model's reliability and its consistent ability to generalize across both outcomes.

Table 15 The CID_SID XGBC ML model classification report.

| | precision | recall | F1-score | support |
|---|---|---|---|---|
| Active (target 1) | 0.81 | 0.94 | 0.87 | 11000 |
| Inactive (target 0) | 0.92 | 0.79 | 0.85 | 11000 |
| | | | | |
| accuracy | | | 0.86 | 22000 |
| macro avg | 0.87 | 0.86 | 0.86 | 22000 |
| Weighted avg | 0.87 | 0.86 | 0.86 | 22000 |

The provided LIME analysis of XGBC based on IUPAC tokenised data offers clear local interpretability for a correctly classified instance, validating the model's overall strong Accuracy of 0.8616. For the instance where the True Label (0) matched the Model Prediction (0), the decision was overwhelmingly driven by the feature SID. Specifically, the condition SID > 49728156.00 contributed a strong positive influence (~ 0.101) towards the prediction of Label 0. In contrast, the feature CID had a negligible, slightly negative influence (~-0.009), indicating that the high value of the SID feature was the primary, almost exclusive reason the XGBC model confidently and correctly assigned this data point to the inactive or negative class.

The XGBC applied to the transformed MORGAN2 dataset obtained an Accuracy of 84.7%, a Precision of 88.1%, a Recall of 86.6%, an F1-score of 87.2%, and an ROC of 84.3%. Tracking for the point where the deviation between train and test accuracy was higher than 5% point

max_depth=6 where the train Accuracy was 90.1% and the test Accuracy was 84.7%. So, the choice was max_depth=5 with train Accuracy of 88.2% and test Accuracy of 83.7.0%. This process, which guided the model's optimization, is visually in Figure 13.

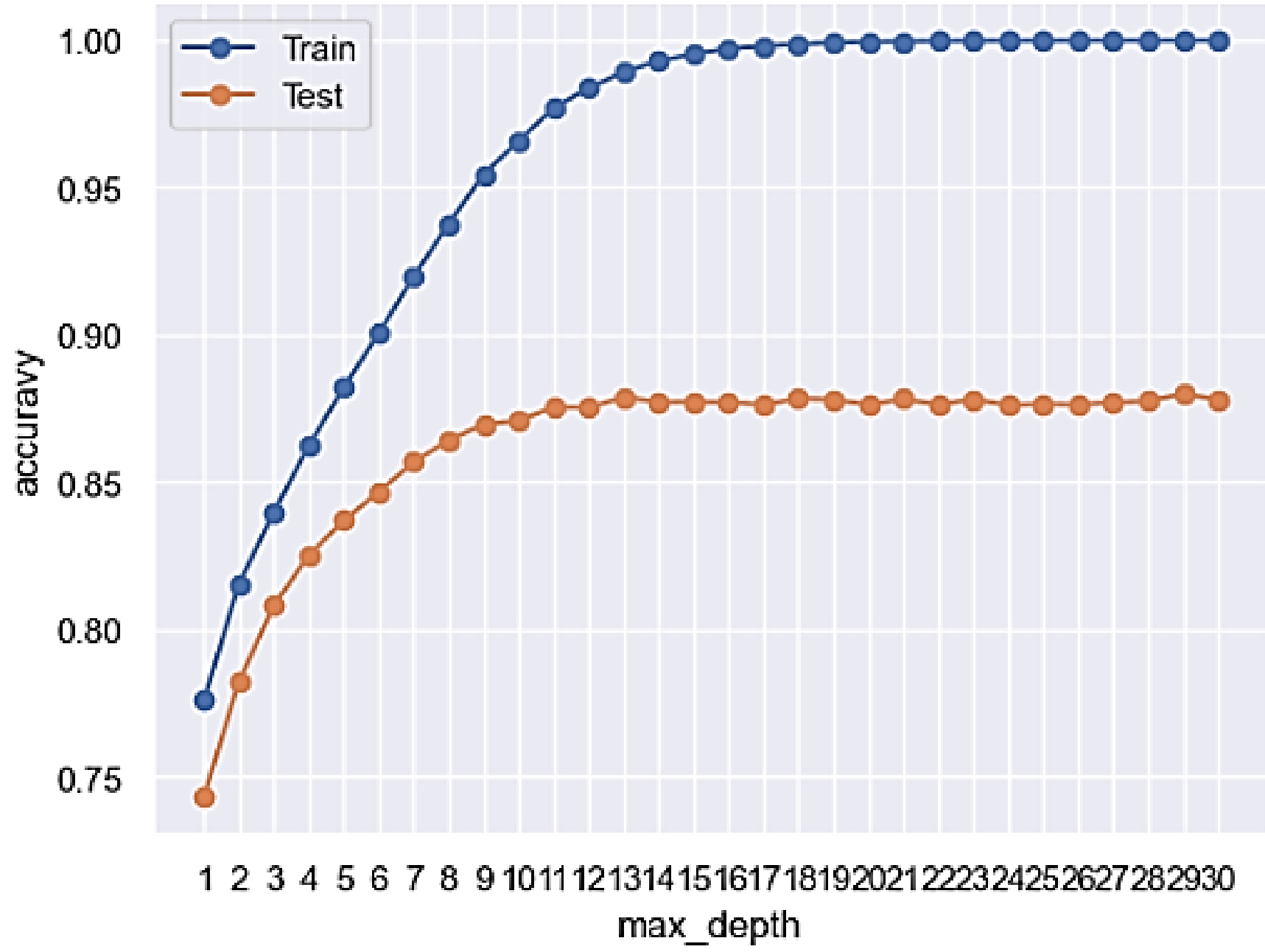

Figure 13.
Overfitting analysis of the XGBC based on transformed MORGAN2 dataset: prediction accuracy vs maximum depth of the decision tree. The blue line is the training accuracy. The orange line is the test accuracy. The deviation between the testing and training accuracy higher than 5% was considered as an indication for early stopping.

A high degree of strong and stable performance was exhibited by the XGBC when it was applied to the transformed MORGAN2 dataset and run with a maximum depth of 5. The model's single-run effectiveness was initially demonstrated by an Accuracy of 83.7%, a high Precision of 87.7%, a Recall of 85%, an F1-score of 86.3%, and an initial ROC of 83.4%. Crucially, the model's reliability and generalizability were confirmed through five-fold CV, where metrics were maintained with exceptionally low variance: the Accuracy was found to be 83.11% ±0.11, the Precision 87.04 ±0.11, the Recall 84.59% ±0.14, and the F1-score 85.8%±0.10. Furthermore, the ROC AUC was observed to be particularly strong at 90.87%± 0.16, suggesting the model possesses excellent and highly consistent discriminative ability across different data subsets. The detailed results are shown in Table 16.

Table 16. Five-fold cross-validation of XGBC based on MORGAN2 transformed features with StratifiedKFold

| **Metrics** | | **Fold** | | | | | **Average across the folds** |
|---|---|---|---|---|---|---|---|
| | | **1** | **2** | **3** | **4** | **5** | |
| **Accuracy [%]** | | 82.97 | 83.30 | 83.10 | 83.10 | 83.06 | 83.11±0.11 |
| **Precision [%]** | | 86.99 | 87.23 | 86.90 | 87.07 | 87.02 | 87.04±0.11 |
| **Recall [%]** | | 84.39 | 84.74 | 84.78 | 84.53 | 84.54 | 84.59±0.14 |
| **F1-Score [%]** | | 85.67 | 85.96 | 85.82 | 85.78 | 85.76 | 85.80±0.10 |
| **ROC AUC [%]** | | 91.00 | 90.99 | 90.84 | 90.96 | 90.58 | 90.87±0.16 |
| **Confusion Matrix** | **TN** | 5224 | 5245 | 5208 | 5230 | 5224 | 5226.2±13.31 |
| | **FP** | 1241 | 1220 | 1257 | 1234 | 1240 | 1238.4±13.35 |
| | **FN** | 1535 | 1501 | 1497 | 1521 | 1520 | 1514.8±15.66 |
| | **TP** | 8298 | 8332 | 8336 | 8312 | 8313 | 8318.2±15.66 |

The hyperparameter tuning using Optuna for the XGBC model was unsuccessful in the five optimisation studies conducted, achieving an accuracy of 83.6% compared to an accuracy of 84.7% when the model was set to default hyperparameters.

The classification performance of the XGBC model, utilizing the MORGAN4 features, was analysed based on the resulting confusion matrix, which is visually represented in Figure 14. Out of all predictions, TP were observed to be 11,368 and TN were 6,942, indicating the number of instances correctly classified as positive and negative, respectively. Misclassifications were also recorded, with FP totalling 1,703 and FN reaching 1,821. Overall, the model's balanced predictive quality, taking all four outcomes into account, was demonstrated by an MCC of 0.6629, signifying a good level of correlation between the true and predicted labels.

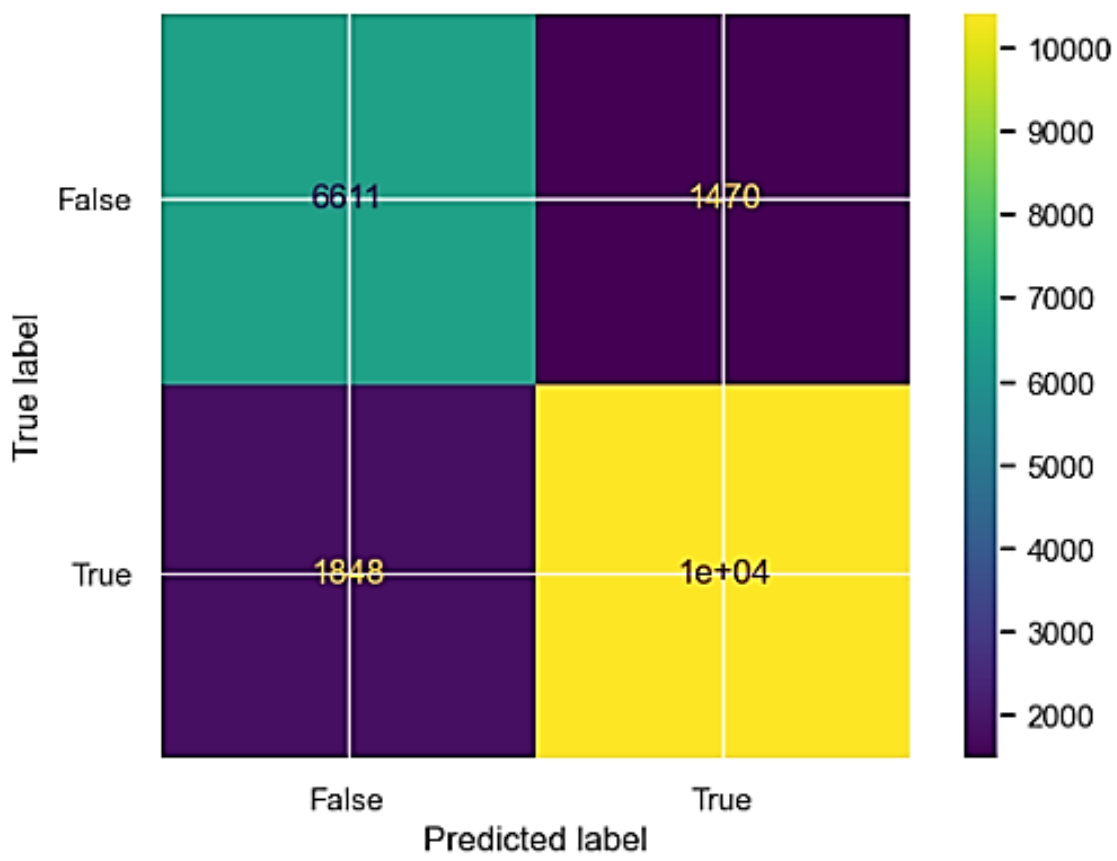

Figure 14. Confusion matrix of the XGBC based on transformed MORGAN2 dataset

The XGBC applied to the transformed by RDKit SMILES dataset obtained an Accuracy of 88.2%, a Precision of 91.7%, a Recall of 88.4%, an F1-score of 90%, and an ROC of 88.1%. Tracking for the point where the deviation between train and test accuracy was higher than 5% point max_depth=8, where the train Accuracy was 95.1% and the test Accuracy was 89.1%. So, the choice was max_depth=7 with train Accuracy 93.7% and test accuracy 88.8.0%. This process, which guided the model's optimisation, is visually shown in Figure 15.

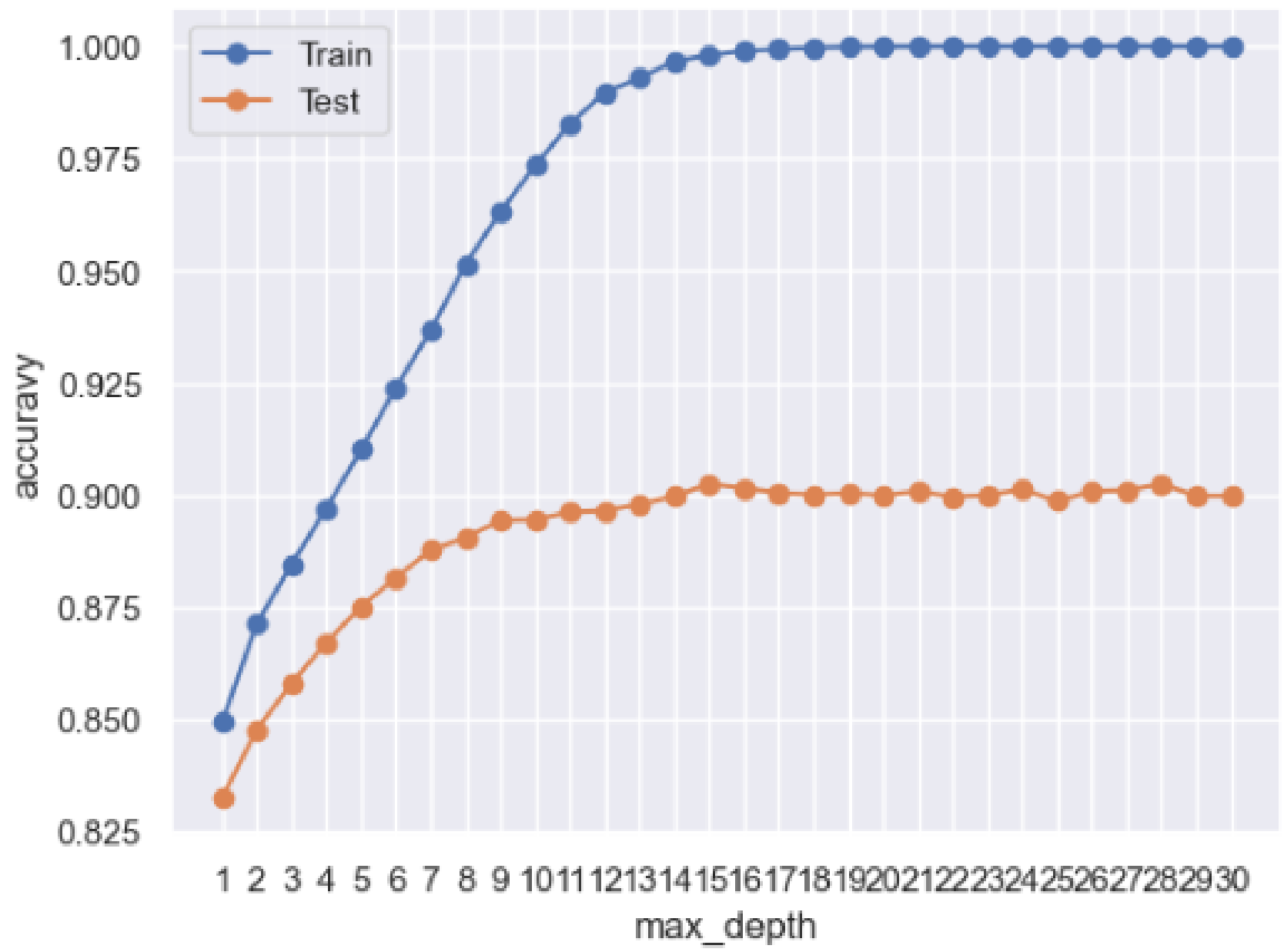

Figure 15.
Overfitting analysis of the XGBC based on dataset with transformed by RDKit SMILES: prediction accuracy vs maximum depth of the decision tree. The blue line is the training accuracy. The orange line is the test accuracy. The deviation between the testing and training accuracy higher than 5% was considered as an indication for early stopping.

The XGBC leveraging RDKit-transformed SMILES features delivered excellent and stable performance, making it a highly reliable candidate for virtual screening. The model achieved a strong single-run test performance with an Accuracy of 88.8% and a particularly high Precision of 91.9%, signifying that most compounds predicted as active will genuinely be active, thus minimising false positives in laboratory follow-up. Furthermore, the 5-fold CV results, consistently clustered around an Accuracy of 88.22% with a minimal standard deviation of ±0.18, confirm the model's high stability and robustness across different data subsets. The outstanding CV ROC of 95.19% with a negligible variance of ±0.09 is especially noteworthy, demonstrating the model's superior ability to discriminate between active and inactive compounds, which is the most critical factor for a practical predictive model in cheminformatics. The detailed results are shown in Table 17.

Table 17. Five-fold cross-validation with StratifiedKFold of XGBC based on RDKit transformed SMILES

| **Metrics** | | **Fold** | | | | | **Average across the folds** |
|---|---|---|---|---|---|---|---|
| | | **1** | **2** | **3** | **4** | **5** | |
| **Accuracy [%]** | | 88.06 | 88.42 | 88.25 | 88.41 | 87.98 | 88.22±0.18 |
| **Precision [%]** | | 91.38 | 91.63 | 91.01 | 91.56 | 91.09 | 91.33±0.25 |
| **Recall [%]** | | 88.57 | 88.93 | 89.35 | 89.00 | 88.76 | 88.92±0.26 |
| **F1-Score [%]** | | 89.95 | 90.26 | 90.17 | 90.26 | 89.91 | 90.11±0.15 |
| **ROC AUC [%]** | | 95.15 | 95.30 | 95.13 | 95.31 | 95.09 | 95.19±0.09 |
| **Confusion Matrix** | **TN** | 5643 | 5666 | 5597 | 5657 | 5610 | 5634.6±26.75 |
| | **FP** | 822 | 799 | 868 | 807 | 854 | 830.0±26.74 |
| | **FN** | 1124 | 1089 | 1047 | 1082 | 1105 | 1089.4±25.66 |
| | **TP** | 8709 | 8744 | 8786 | 8751 | 8728 | 8743.6±25.66 |

Hyperparameter tuning using Optuna for the XGBC model was unsuccessful in the five optimisation studies conducted, achieving an accuracy of 87% compared to an accuracy of 88.8% when the model was set to default hyperparameters.

The high-performing XGBC model, based on the dataset transformed by RDKit SMILES, achieved a strong, balanced outcome, as evidenced by an MCC of 0.7678. This high MCC confirms the model's reliability across both classes. The confusion matrix further details this success: out of the positive predictions, 10,966 were TP, while only 962 were FP. This low FP count is a critical advantage in virtual screening, minimising the cost of testing inactive compounds. On the negative side, 7,119 instances were correctly identified as TN, with 1,325 cases missed as FN. Overall, the metrics confirm that the XGBC is a robust classifier, demonstrating a high and well-balanced predictive capability for both active and inactive compounds. The confusion matrix of the XGBC based on the dataset transformed by RDKit SMILES is visually represented in Figure 16.

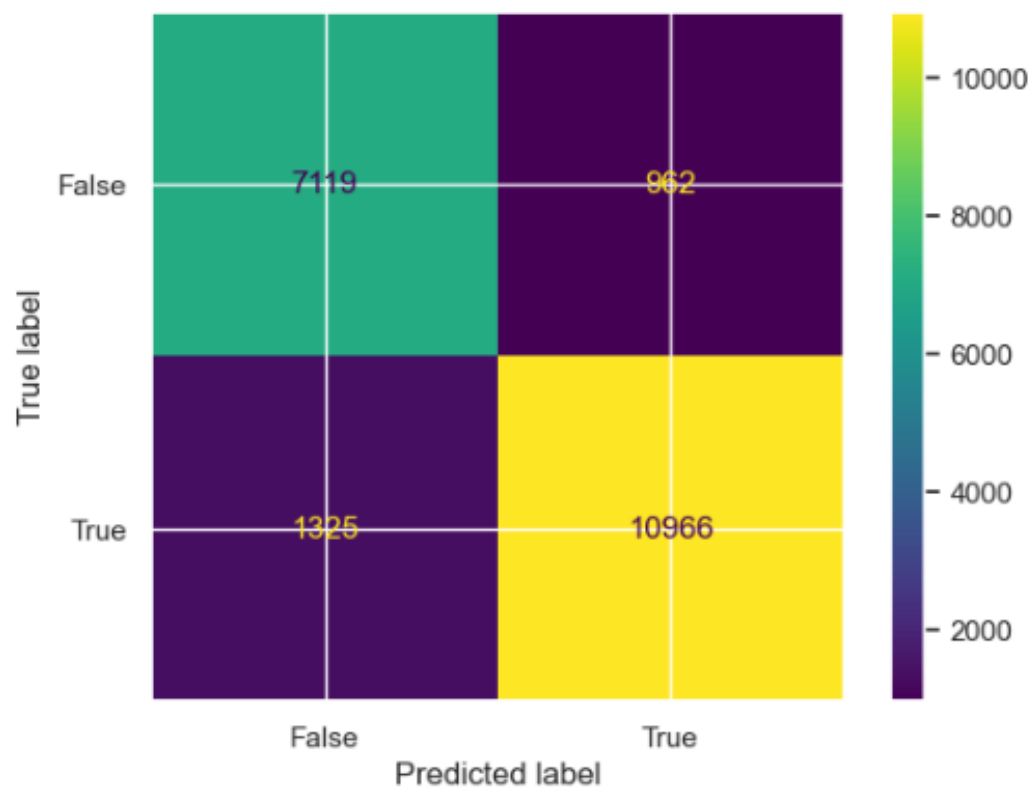

Figure 16. Confusion matrix of the XGBC based on dataset with transformed by RDKit SMILES

The comparative analysis of XGBC models, assessed using the MCC, reveals that the choice of molecular descriptors dictates predictive performance for this classification task. The model utilizing features derived from SMILES RDKit achieved the highest MCC of 0.7912, indicating the most accurate and balanced classification performance. This result is significantly superior to the other two feature sets. The model based on CID_SID features achieved the second-best performance with an MCC of 0.7311. The least effective model used MORGAN2 circular fingerprints, yielding the lowest MCC of 0.6605. This ranking suggests that the specific.

## Conclusion

The proposed methodologies are expected to be widely applicable to any case study featuring a bioassay with a significant number of labelled records. Although both the CID_SID model and the ML model based on IUPAC names predict the same specific functionality, TDP1 inhibition, their implementations and utility for biochemical research are distinct. The CID_SID ML model can be integrated in a time- and cost-efficient suite of ML models, predicting the functionalities of compounds beyond their primarily designed purpose. Conversely, the ML model based on the IUPAC data is used to generate a descending order of feature importance of RFC. While the initial feature importance ranking from the RFC provides a preliminary computational (*in silico*) assessment, translating this hierarchy into reliable, real-world insights necessitates statistical re-evaluation and reordering through rigorous approaches. Ultimately, confirming these statistically prioritised features in the laboratory would be a major benefit to early drug discovery. It would quickly guide researchers toward the most functionally relevant groups, boosting the speed and efficiency of human intelligence-driven research.

*Scientific contribution*

- By tokenising IUPAC names, the proposed methodology established a reliable, AI hallucination-free foundation for analysis that surpasses the low ML scores of IUPAC token-based RFC models. The result is the production of focused lists identifying key functional groups for TDP1 inhibition, thereby providing clear direction and accelerating drug discovery by human researchers.
- Development of CID_SID ML models, thereby increasing the number of CID_SID ML models that can be integrated into a cost- and time-efficient framework predicting functionalities of small biomolecules other than their original purpose.

*Limitations*

- Parsing IUPAC names generates thousands of features, necessitating a substantial dataset for both ML model training and effective feature list generation. This large dataset is critical to satisfy the practical ML guideline, suggesting that the number of data rows should be at least ten times greater than the number of columns (features). Techniques such as HTS are necessary to provide the volume of labelled data required.
- While the importance of cross-referencing predicted active fragments with published SAR data for establishing consistency with known medicinal chemistry principles is fully acknowledged, a comprehensive, systematic review of all historical SAR is unfortunately constrained by resource limitations and the lack of access to the necessary proprietary literature databases and advanced cheminformatics tools.
- External validation is a critical step; however, reliance must currently be placed on rigorous internal validation procedures because a suitable, independently collected external dataset has not yet been collected or made available for this purpose
- Because the exact coordinates and tolerance values of the proprietary TDP1 pharmacophore model are not publicly available, the top-ranked functional groups could not be checked against known pharmacophores for TDP1 inhibitors. Access to this information is restricted to licensed software, such as MOE or LigandScout, which the authors of this article were unable to obtain.
- Lack of laboratory confirmation of the computational obtained results, which means the computational results are currently hypothetical and unproven in a real biological system.
- Hyperparameter tuning of RFC with Optuna and Boruta feature importance was performed with only five studies/iterations.
- Manual calculation of the mean Ratio of Active and Inactive cases for the functional group/fragment having a different spelling.
- Only a few samples of chemical compounds related to the study were fully screened for PAINs.

*Future directions*

- Limited by an ongoing study, the authors will not continue to further mature the methods presented here. However, a valuable next step for advancing biochemical research is to integrate the existing CID_SID ML model, develop more models like it, and combine them into a single PoC framework. This framework would then allow for an estimation of the time benefits it provides compared to using features transformed by tools like MORGAN2 or SMIELS/RDKit. The CID_SID machine learning model, despite exhibiting a lower MCC (0.7311) than the SMILES RDKit model (0.7912), offers a significant practical advantage: it eliminates the need for on-the-fly descriptor calculations. Since the features associated with CID/SID are pre-computed and readily available through PubChem, the substantial time investment required for generating transformations like those necessary for SMILES RDKit is avoided. While this time saving may be negligible for a single model run, it is expected to become highly significant when scaling up to a large suite of integrated ML models used for predicting new functionalities for vast sets of test compounds. The quantification of the time saved by utilizing the CID_SID ML approach is designated as an objective for future investigation.

- Despite the justification provided in the paper for keeping PAINs, a comprehensive exploration of all proposed methodologies using datasets from which PAINs have been removed would be an appealing avenue for future investigation. It can be expected that the fidelity of the resulting models and analyses would be significantly increased by this step. By eliminating these known interference compounds, the computational results, including SAR analysis and ML predictions, will be based on molecules whose activity is more likely attributed to a specific biological interaction rather than an assay artefact, thereby yielding more reliable and chemically meaningful insights for drug discovery.
- A promising variant for future exploration that aims to significantly facilitate biochemical research involves using the PubChem Substructure Fingerprint instead of IUPAC names as molecular descriptors. In this approach, the dataset features would be the 881 bits of the PubChem fingerprint, each corresponding to the presence or absence of a specific structural motif, while the labels would be derived from an HTS bioassay of interest. Generating a descendant ordered list of feature importance using RFC for these 881 bits, and subsequently processing this list further using methods such as Fisher's Exact Test, or determining the most and least desirable functional groups based on the relative proportion of active cases, would yield highly actionable information for biochemical researchers. This structural guidance would streamline their synthetic work by identifying the most potent fragments for activity, thereby speeding up research, lowering costs, and accelerating drug discovery.
- Running the hyperparameter tuning study only five times is insufficient and should be increased to at least 100 iterations. Increasing the number of study runs significantly improves the probability of finding a globally optimal or near-optimal hyperparameter configuration, which is necessary to maximise model performance and ensure the final results are not due to chance, thereby increasing confidence in the tuning process.
- To ensure Boruta suggests features despite an initial lack of confirmation, several tuning actions must be considered. Better statistical convergence can be allowed by increasing the maxRuns, or potential feature interactions can be better captured by adjusting the max_depth of the underlying RFC estimator. Furthermore, the root of the problem, a constant or near-constant target variable, or highly correlated features masking true importance, must be addressed through a thorough examination of the dataset, which would solve the troubleshooting requirement.
- Automation of merging the same functional groups/fragments that have different spellings during the IUPAC tokenisation
- Analysing the full list with results based on PAIN flags and concluding further investigations

- A comparison of the results obtained by the presented approach and Bio5T+, confirmed in a chemical laboratory, would provide clarification regarding the level of credibility and preferences between both.

**Data and Code Availability Statement**

The Python code is available on GitHub as Jupyter notebook files: https://github.com/articlesmli/IUPAC_ML_model_TDP1

The datasets used for the study are available on Hugging Face:

- Dataset for the IUPAC-based cases, such as most and least desired functional groups, RFC and its feature importance, focused on TDP1 inhibitors [29] https://huggingface.co/datasets/ivanovaml/TDP1_targetsInhibitors_CID_SID_IUPACs_functionalGroups/blob/main/README.md
- Dataset for the CID_SID ML model [50] https://huggingface.co/datasets/ivanovaml/TDP1_targetInhibitors_CID_SID
- PAIN flagged dataset with TDP1 inhibitors, their CID, SIDs and SMILES [45] https://huggingface.co/datasets/ivanovaml/TDP1_inhibitors_no_PAINsn
- Dataset with CID, SID, IUPAC and PAIN [46] https://huggingface.co/datasets/ivanovaml/CID_SID_IUPAC_PAIN

The raw data was provided by PubChem https://pubchem.ncbi.nlm.nih.gov/bioassay/686978

## Conflicts of Interest

The authors declare no conflict of interest